\documentclass[twocolumn]{aastex631}

\usepackage{natbib}
\usepackage{epsfig}
\usepackage{url}

\usepackage{bm}
\usepackage{color}
\begin{document}

\title{Efficiency of Non-Thermal Pulsed  Emission from Eight  MeV Pulsars }

\author{J. Takata}
\affiliation{Department of Astronomy, School of Physics, Huazhong University of Science and Technology, Wuhan 430074, People's Republic of China}
\email{takata@hust.edu.cn}
\author{H.-H. Wang}
\affiliation{School of Physics and Astronomy, Sun Yat-sen University, Zhuhai 519000, People's Republic of China}
\author{L.C.-C., Lin}
\affiliation{Department of Physics, National Cheng Kung University, Tainan 701401, Taiwan}
\author{S. Kisaka}
\affiliation{Physics Program, Graduate School of Advanced Science and Engineering, Hiroshima University, Higashi-Hiroshima, 739-8526, Japan}

\begin{abstract}
We report on the properties of pulsed X-ray emission from eight  MeV pulsars using   XMM-Newton, NICER, NuSTAR and HXMT data. For the five among eight MeV pulsars, the X-ray spectra can be fitted by a broken-power law model with  a break energy of $\sim5-10$~keV. The photon index below and above break energy are $\sim 1$ and $\sim 1.5$, respectively. In comparison with the X-ray emission of the $Fermi$-LAT  pulsars, the MeV pulsars have  a harder spectrum and a higher radiation efficiency in 0.3-10~keV energy bands. By assuming the isotropic emission, the emission efficiency in the keV-MeV bands is estimated to be $\eta_{MeV}\sim 0.01-0.1$, and it is similar to  the efficiency of GeV emission of the $Fermi$-LAT pulsars that have similar spin-down power.  To explain the observed efficiency of the MeV pulsars, we estimate the required pair multiplicity as  $10^{4-7}$ that depends on the emission process (curvature radiation or synchrotron radiation) and the location in the magnetosphere. The large multiplicity indicates that the secondary  pairs that are created by a pair-creation process of the GeV photons produce the X-ray/soft gamma-ray emissions of the MeV pulsars. We speculate that the difference between the MeV pulsars and $Fermi$-LAT pulsars is attributed to the difference in viewing angle measured from the spin-axis, if the emission originates from a  region inside the light cylinder (canonical gap model) or the difference in the inclination angle of the magnetic axis, if the emission is produced  from equatorial current sheet outside the light cylinder.
\end{abstract}
\section{Introduction}

Since the launch of $Fermi$-Large Area Telescope (hereafter $Fermi$-LAT) in 2008, more than 300 gamma-ray emitting rotation-powered
pulsars have been identified~\citep{2023ApJ...958..191S}. The $Fermi$-LAT observations have revealed that the typical spectral energy distribution  of
the energetic pulsars shows  a peak in the GeV bands. Moreover,  the GeV/TeV   observations 
also show  that the observed pulsed  gamma-rays from the pulsars
are produced in the outer magnetosphere, because the gamma-ray photons produced near the stellar surface
cannot avoid the pair-creation process with the strong magnetic field~\citep{2008Sci...322.1221A,2020A&A...643L..14M}. 
The  discovery of $\sim 20$~TeV pulsed  emission from the Vela pulsar also supports this hypothesis~\citep{2023NatAs.tmp..228H}. Either canonical gap model~\citep{1986ApJ...300..500C, 1981IAUS...95...69A} or the equatorial current sheet
model~\citep{1996A&A...311..172L, 2015MNRAS.448..606C,2023ApJ...959..122C} has been considered  as the   possible acceleration region
in the outer magnetosphere. 

The so-called MeV pulsars form  a distinct group from the $Fermi$-LAT pulsars, based on their observed emission
properties~\citep{2014MNRAS.445..604W,2015MNRAS.449.3827K,2017ifs..confE...6H, 2018MNRAS.475.1238K}.
PSR~B1509-58 located in  the supernova remnant MSH~15-52 has been known as an energetic young pulsar since the high-energy observation in the Energetic Gamma-Ray Experiment Telescope  era
\citep{1999ApJ...516..297T}. The $Fermi$-LAT observation confirms the pulsed emission from PSR~B1509-58 in 0.1-1~GeV bands, but the  spectrum from X-ray to GeV bands has a peaking at $\sim 1-10$~MeV, which is about three orders of magnitude smaller than that of the typical $Fermi$-LAT pulsars~\citep{1999A&A...351..119K,2010ApJ...714..927A}. \cite{2018MNRAS.475.1238K} report the $Fermi$-LAT detection of PSR~J1846-0258 and reveal that the spectral shape 
in the X-ray to GeV bands is similar to that of PSR~B1509-58.  Based on the spectral properties reported in
\cite{2014MNRAS.445..604W} and \cite{2015MNRAS.449.3827K}, we select
the eight sources  as the MeV pulsars. Table~1 summaries the spin-period $(P_s)$, dipole magnetic field $(B_s)$, the spin-down power $(L_{sd})$, and the distance  $(d)$ of the eight pulsars. 

It has been revealed that the X-ray emission
of the MeV pulsars has a distinct property from that of the $Fermi$-LAT pulsars. The X-ray emission of the $Fermi$-LAT pulsars is in general  described by the thermal component from the neutron star surface  and/or the non-thermal 
component that has a photon index of $\Gamma_1\sim 1.5-2$ \citep{2013ApJS..208...17A, 2021MNRAS.502..390H}. It has been discussed that the non-thermal X-ray emission is produced by the synchrotron radiation from the secondary electron/positron pairs that are produced by the pair-creation process of the GeV gamma-rays~\citep{2017ApJ...837...76K}.  For the MeV pulsars, on the other hand, the previous X-ray studies have collected evidences that the broadband X-ray spectrum is described by a single power law with a photon index of $\Gamma_1<1.5$ or a power law  with a  turnover in $1-10$~keV  bands [see e.g.  \cite{2016ApJ...817...93C}, \cite{2017ApJ...838..156H} and references therein for PSR~B1509-58, \cite{2009ApJ...690..891K} for PSR J1617-5055,  \cite{2020ApJ...889...23M} for PSR~J1811-1925, \cite{2015MNRAS.449.3827K} for PSR J1813-1749,  \cite{2009MNRAS.400..168L} and references therein for PSR~J1838-0655, \cite{2021ApJ...908..212G} for PSR~J1846-0258, \cite{2011ApJ...729L..16G} and~\cite{2024ApJ...960...78K} for PSR~J1849-0001, and~\cite{2007ApJ...663..315L} for PSR~J1930+1852].  We also refer  \cite{2015MNRAS.449.3827K} for a comprehensive observational study of the hard X-ray/soft gamma-ray emissions of the young pulsars. Although several emission models  have been proposed~\citep{2013ApJ...764...51W, 2014MNRAS.445..604W, 2017ifs..confE...6H}, the emission process and even the emission region of the MeV pulsars are still not clearly understood.

In this paper, we revisit the X-ray analysis for  the eight MeV pulsars using the data taken by
XMM-Newton telescope, the Neutron Star Interior Composition Explorer (NICER),  Nuclear Spectroscopic Telescope Array (NuSTAR), and Hard X-ray Modulation Telescope (HXMT).
Main purpose of this study  is (i)  to estimate the efficiency of the non-thermal  emission of the MeV pulsars ($\eta_{MeV}$),
which is the ratio of the luminosity in X-ray/soft gamma-ray bands  to the spin-down power, and (ii) to compare it with the efficiency of the ``GeV emission'' ($\eta_{GeV}$)  of the $Fermi$-LAT pulsars.  The emission
efficiency can be used to  constrain  the particle acceleration and emission processes. For the $Fermi$-LAT pulsars, for example,  the efficiency of the GeV emission  is correlated with the spin-down power ($L_{sd}$) as  $\eta_{GeV}\propto L^{-1/2}_{sd},$ and the origin of the correlation has been  investigated with the emission model~\citep{2010ApJ...715.1318T,2019ApJ...883L...4K}.  The emission efficiency can be linked with so-called  multiplicity ($\kappa$) that is the ratio of the number density of the emitting electrons/positrons  over the Goldreich-Julian value,
$n_{GJ}=\Omega_s B/(2\pi ce)$, where  $\Omega_s$ is the spin angular frequency, $B$ is the local magnetic field strength, $c$ is the speed of light and $e$ is the elementary charge \citep{1969ApJ...157..869G}.  The required multiplicity to explain the observed efficiency can be used to investigate the emission process, too.

The paper is organized  as follows. In section~\ref{data}, we describe the data reduction of each telescope.  In section~\ref{result}, we  estimate the efficiency of the non-thermal emission of the MeV pulsars by fitting the observed spectra. In section~\ref{discussion}, we compare the efficiency of the MeV pulsars with that of the $Fermi$-LAT pulsars  and  we discuss the required  multiplicity for  different  emission processes. Then we summarize our results in section~\ref{summary}.  In this paper, we apply  Gaussian-cgs units system to express the physical quantities.

\begin{deluxetable}{ccccc}
  \tablecolumns{8}
  \tabletypesize{\footnotesize}
  \tablecaption{Spin-down parameters of the eight MeV pulsars}
  \tablehead{
    \colhead{PSR}  &
    \colhead{$P_s$~(ms) }&
    \colhead{$B_s$~($10^{12}$G)} &
    \colhead{$L_{sd}~(10^{36}{\rm erg~s^{-1}})$} &
    \colhead{$d$ (kpc)}
  }
  \startdata
  B1509-58 & 152  & 15 & 17 & 4.4 \\
 J1617-5055 & 69.4  & 3.1 & 16 & 4.7 \\
  J1811-1925  & 64.7 & 1.7&  6.4& 5.0  \\
  J1813-1749  & 44.7  &2.4  & 56 & 6.1 \\
  J1838-0655 & 70.5  & 1.9& 5.5& 6.6 \\
  J1846-0258 & 327  & 49& 8.1& 5.8\\
  J1849-0001 &  38.5 & 0.75 & 9.8 & 7.0$^{a}$ \\
  J1930+1852 & 137 & 10& 12 &  7.0
  \enddata
  \tablenotetext{}{Values of the  parameters are obtained from ATNF Pulsar Catalog~\citep{2005AJ....129.1993M}.}
 \tablenotetext{\rm a}{Distance assumed in \cite{2011ApJ...729L..16G}.}
\end{deluxetable}

\section{Data reduction}
\label{data}
We download archival  X-ray data from High Energy Astrophysics Science Archive Research  Center\footnote{\url{https://heasarc.gsfc.nasa.gov/cgi-bin/W3Browse/w3browse.pl}} and  XMM-Newton Science Archive\footnote{\url{http://nxsa.esac.esa.int/nxsa-web/}}. For PSR~B1509-58,  we also analyze the  HXMT data, which is downloaded from the HXMT web\footnote{\url{http://hxmten.ihep.ac.cn/}}. We use HEASoft version  6.32.1, FTOOL~(Nasa High Energy Astrophysics Science Archive Research Center (Heasarc), 2014),  XMM-Newton SAS version 21.0.0, and HXMTDAS. The information of the observations is summarized in Table~A1-A3.

\subsection{XMM-Newton}
Because the MeV pulsars are usually located in their pulsar wind nebulae, it is required to carry out
a spin-phase resolved spectroscopy to extract the pulsed spectrum. For the XMM-Newton, therefore, we
analyze the data taken by PN-CCD (hereafter PN),  because the timing resolution data taken by MOS-CCD is
insufficient for the MeV pulsars.  We use the SAS task\footnote{\url{https://www.cosmos.esa.int/web/xmm-newton/sas-threads}} \verb|epproc|
to obtain the cleaned event file and then we apply the task \verb|barycen|
to carry out barycentric correction for the arrival time of the each event (ephemeris DE 405).  We remove epochs with background flaring by limiting the count rate above 10 keV to $<0.4~{\rm count~s^{-1}}$.

For each event file analyzed in this
study, we create  $Z^2_1$-periodogram~\citep{1983A&A...128..245B} and  identify the periodic signal at the spin-frequency of the pulsar.
We fold each event file using the best frequency in each periodogram,  and create the light curve to determine the on-pulse and off-pulse phases. For all pulsars in this study, the folded light curve is described by a single broad peak, such as  the pulse profile of PSR~B1509-58 in Figure~\ref{b1509-l}  (Figure~\ref{pulse} for eight MeV pulsars).
We therefore define the on-pulse phase covering
60\% of the single spin phase and
the off-pulse phase with other 40\%, as indicated in Figures~\ref{b1509-l} and~\ref{pulse}.  We create the spectra of the on-pulse phase and the off-pulse phase, and obtained the spectrum of the pulsed component by subtracting the spectrum of the off-pulse phase from the spectrum of the on-pulse phase. For XMM-Newton data, we group the source spectrum such that each spectral bin has  a minimum signal-to-noise ration of 3. We use the standard tasks of the SAS to generate the  spectral file, response matrix file (rmf) and auxiliary response file (arf). We confirm with SAS command \verb|epatplot|\footnote{\url{https://www.cosmos.esa.int/web/xmm-newton/sas-thread-epatplot}} that the effect of the pile-up on the spectrum is negligible for the eight MeV pulsars.

\subsection{NICER}
NICER observed our targets   except for PSRs~J1617-5055 and J1930+1852. We apply  \verb|nicerl2| to each data set to obtain the cleaned event file and we apply a
barycentric time correction to the created event  file using the \verb|barycorr| task of \verb|HEASoft|. NICER repeatedly observes one target and the exposure of each observation is usually  several kilo-seconds.
For each pulsar, we create a local ephemeris  by combining several data sets; all ephemerides and obs. IDs
of the NICER data used in this study are summarized in Table~A3. We also refer the ephemerides reported in \cite{2020MNRAS.498.4396H} for J1831-1749 and \cite{2023ApJ...952..120H} for J1846-0258 

We merge the multiple observations using 
the task \verb|niobsmerge|\footnote{\url{https://heasarc.gsfc.nasa.gov/docs/nicer/analysis_threads/combine-obs/}}. We create the good time interval files for on-pulse and off-pulse phases, and extract the events using \verb|extractor|. We use \verb|niextspect| to extract the spectrum from the event files,  and  create  the  rmf  and arf from  the merged event file using the task \verb|nicerl3-spect|.  We group the events so that each spectral bin contains a count of $>1000$, since each channel before binning records a few
  hundred events. We checked the different threshold of the photon counts for  the binning and confirmed
that results of the fitting is insensitive to  how we  bin the data.

\subsection{NuSTAR}
First, we carry out the tasks \verb|nupipeline| and \verb|nuproducts| of HEASoft to obtain the clean event list  after
barycentric time correction. In $Z^2_1$-periodogram, we can confirm the strong spin signal in the data extracted from the source region.  We fold each data set with the best
frequency in  $Z^2_1$-periodogram and create the good time interval files for on-pulse phase and off-pulse phase.
We rerun to the task \verb|nuproducts| with the option \verb|usrgtifile|  and obtain the spectra of the on-pulse/off-pulse phases and response files (rmf and arf).  Using the pipeline  \verb|nuproducts|, we group the on-pulsed spectra to ensure at least 100 counts per spectral bin, since each channel before binning records a few ten events.

\subsection{HXMT}
\begin{figure}
  \epsscale{1}
  \centerline{
    \includegraphics[scale=0.6]{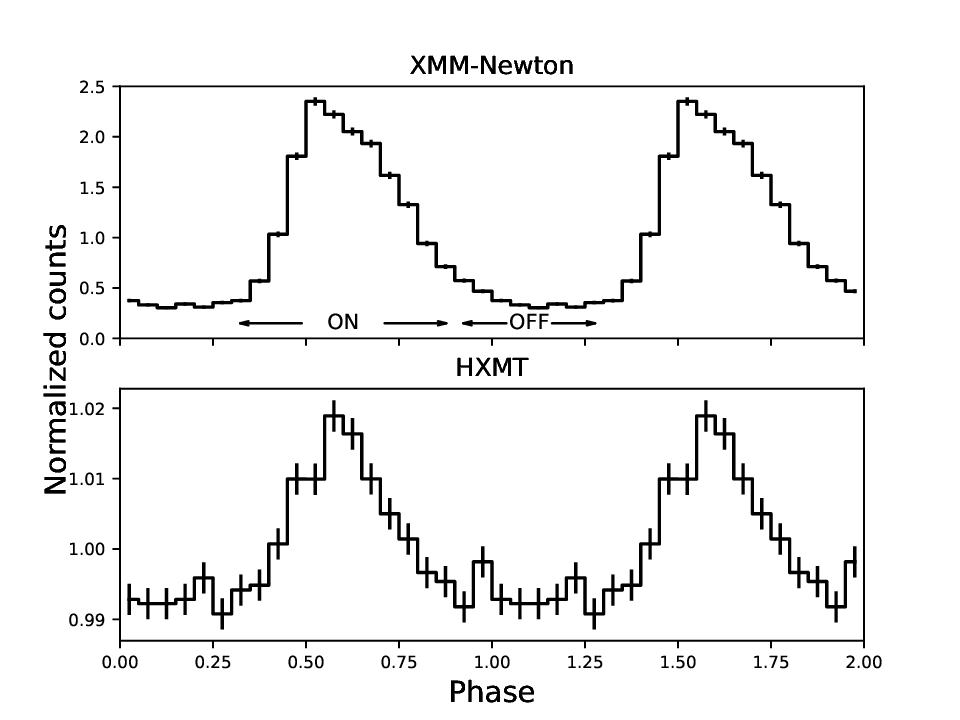}}
  \caption{Pulse profiles of PSR~1509-58. Upper and lower panels show the data taken by XMM-Newton PN  and HXMT HE cameras, respectively.}
  \label{b1509-l}
\end{figure}

 HXMT has three  payloads [high-energy (20-250~keV), medium-energy (8-35~keV) and low-energy (1-12~keV) X-ray telescopes] and can cover a wide energy range in X-ray bands.
In this study, we use the  data of the high-energy X-ray telescope of  PSR~B1509-58 taken in 2017 July (Obs. ID. P0101324001);  we cannot find a significant pulsation in the data taken by medium- and low-energy telescopes. We apply  the task \verb|hepical| to remove spike events caused by electronic systems
and \verb|hegtigen|  to select good time interval file  with the conditions that  (i)
the pointing offset angle is smaller than $<0.04^{\circ}$,  (ii) the pointing direction is  $>1^{\circ}$
above Earth, (iii) the geomagnetic cut-off rigidity is bigger than $>8$~GeV,
and (iv) the South Atlantic Anomaly does  not occur within 300~s.
We use  \verb|hxbary| for the barycentric correction of the HXMT data, and we confirm a significant  periodic signal in the data of the high-energy X-ray telescopes (Figure~\ref{b1509-l}). We use the \verb|hspec-merge| to merge the spectra and the response files extracted from the different data sets. For HXMT data, we group the on-pulsed spectra to ensure at least  1000 counts per spectral bin.

\section{Results}
\label{result}
In this section, we present the results of the spectral fitting for the pulsed components of the eight MeV pulsars.
We carry out a joint fit of the spectra  taken by the four telescopes. Using \verb|Xspec|,
we fit the spectra with a single power law   model \verb|constant*tbabs*powerlaw| (hereafter s-pl model) or a broken power law model  \verb|constant*tbabs*bknpower| (hereafter bk-pl model), where \verb|constant| accounts for
cross-calibration mismatch among the telescopes.  In previous studies \citep{2001A&A...375..397C, 2016ApJ...817...93C, 2017ApJ...838..156H}, a
 log-parabola  (\verb|logpar|) model, which is described by a function of $F(E)\propto E^{-[a+b{\rm log}(E/E_0)]}$ with $a$ and $b$ being  fitting parameters  and $E_0$ the energy scale, is used to describe the curved X-ray  spectrum of the MeV pulsars. In this study, the bk-pl model is preferred over  the logpar model, because it is  more intuitive to physically interpret. As we discussed in the following section~\ref{syncrad}, for example, we will use the fitting break energy of bk-pl model to constrain the multiplicity of the pairs.

\begin{figure*}
  \epsscale{1}
  \centerline{
    \includegraphics[scale=0.6]{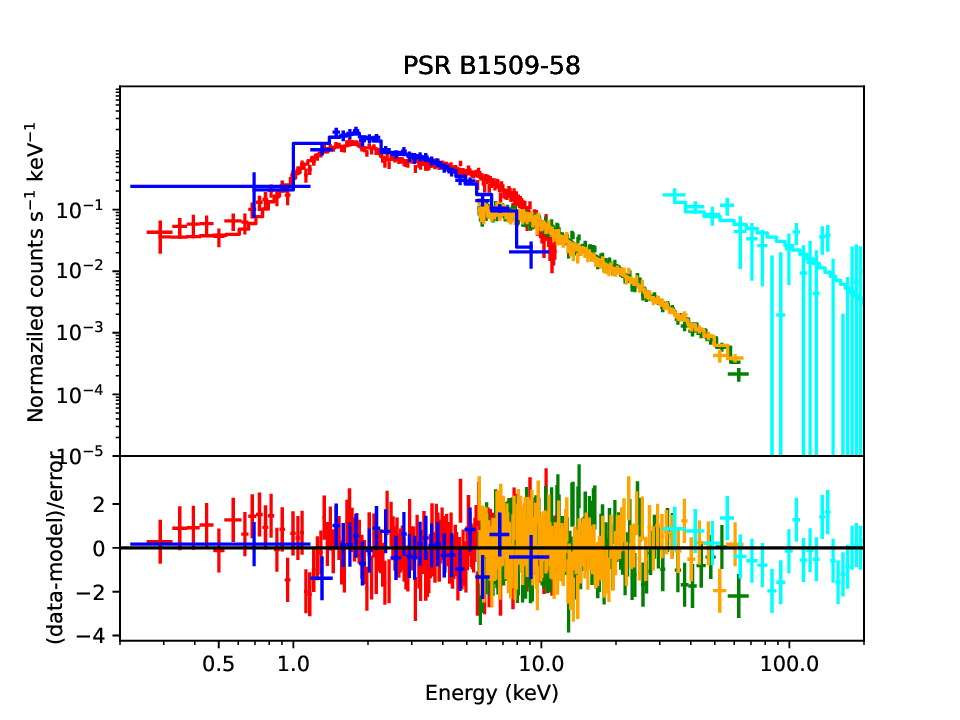}
    \includegraphics[scale=0.6]{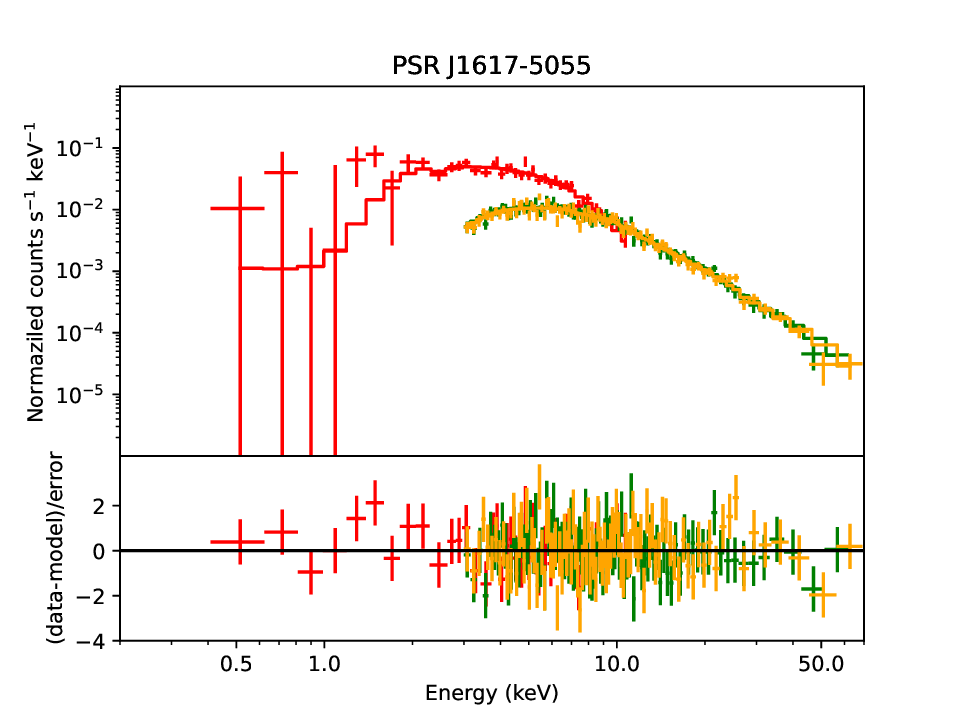}}
  \caption{X-ray broad bands spectra for PSRs B1509-58 (left) and J1617-5055 (right). The red, blue, green, orange and light blue colors correspond to the data taken by
  XMM-Newton PN, NICER, NuSTAR FPMA, FPMB and HXMT data, respectively.}
  \label{b1509}
\end{figure*}

In our analysis, we fix the \verb|constant| factor
for the XMM-Newton observation at unity.
Figures~\ref{b1509} and \ref{j1838}-\ref{j1930} show the pulsed spectra for the eight pulsars.
 Although we choose only  one spectrum of each detector in the figure for the representation,  we use more available data (as shown in Table~A1-A3) for the spectral fitting and fit all spectra simultaneously; the Figures~\ref{b1509} and \ref{j1838}-\ref{j1930} represent the spectra taken by the earlier observation if there are multiple observations. Table~2 summarizes the results of the fitting  spectral parameter, the estimated isotropic
luminosity, and its emission efficiency in 0.3-150~keV bands.

\begin{figure*}
  \epsscale{1}
  \centerline{
    \includegraphics[scale=0.5]{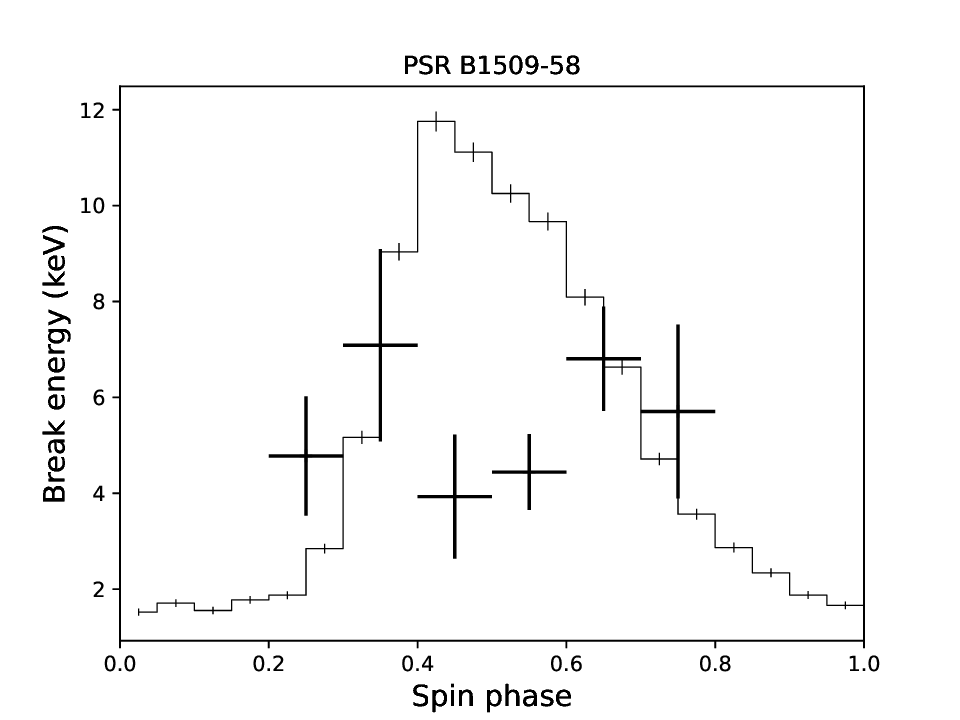}
    \includegraphics[scale=0.5]{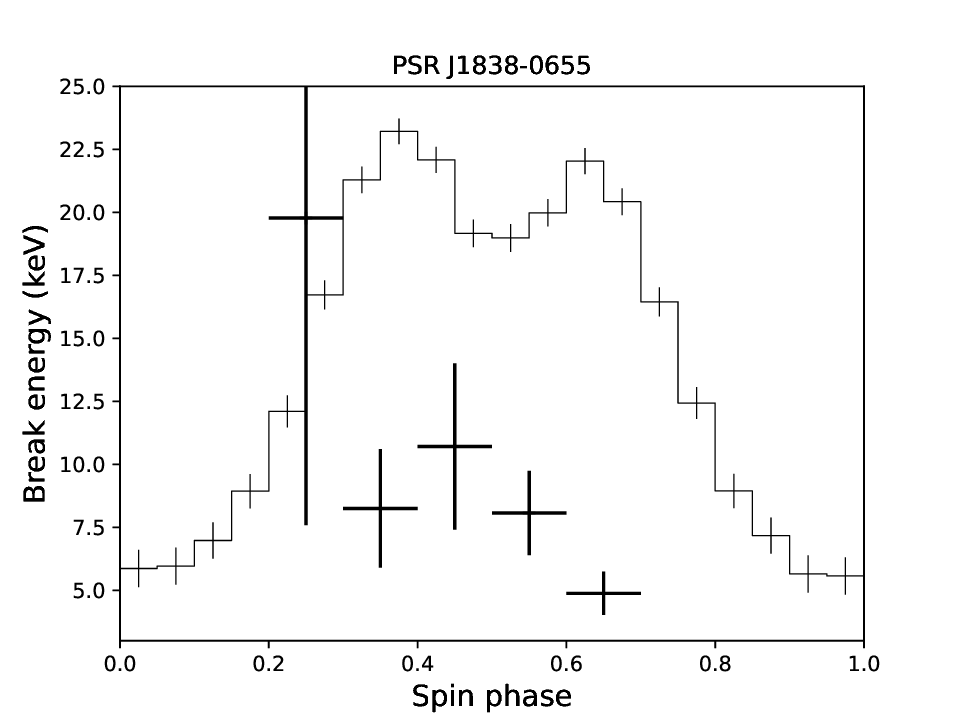}}
  \caption{Evolution of the break energy ($E_b$) with the spin phase for PSR~B1509-58 (left) and for PSR~J1838-0655 (right).
    Histogram show the pulse profiles measured by XMM-Newton PN. Hydrogen column density is fixed to
    $N_H=0.9\times 10^{22}~{\rm cm^{-2}}$ for PSR B1509-58 and $6.0\times 10^{22}~{\rm cm^{-2}}$ for PSR~J1838-0655. The break energy at the wings of the pulse
    ($\sim 0.25$ and $\sim 0.75$ spin-phase) of  PSR~J1838-0655 cannot
    be constrained. }
  \label{b1509-p}
\end{figure*}

\subsection{PSR B1509-58}
For PSR B1509-58, there is one archival XMM-Newton data taken with the  PN camera under  the small widow mode, whose timing resolution ($\sim 5.7$~ms\footnote{\url{https://heasarc.gsfc.nasa.gov/docs/xmm/uhb/epicmode.html}}) is sufficient enough to carry out the phase-resolved spectroscopy. There are four data sets obtained by NuSTAR observations. NICER observed the target in 2017 July and August, for which we create local  ephemerides  to  fold the data. The left panel of Figure~\ref{b1509} shows the extracted spectra of the pulsed component. 

First we fit the spectrum with a  s-pl model and obtain the hydrogen column density of $N_H=1.32(5)\times 10^{22}~{\rm cm^{-2}}$ and the photon index of $\Gamma_1=1.41(1)$ with $\chi^2/D.O.F.=2023.99/2023$ (Table~2).
We find that the predicted $N_H$ is higher than $N_H\sim0.9\times 10^{22}~{\rm cm^{-2}}$ reported in the previous studies using Bepposax observation \citep{2001A&A...375..397C, 2023ApJ...957...23R} (the previous studies used \verb|Phabs| for the absorption model in \verb|Xspec|, but we confirm that the fitting results are not sensitive to the absorption model and obtain a similar to current value).

Second, we fit the spectra with a  bk-pl model. As Table~2 indicates,
the improvement of the fitting from the s-pl model is significant
with an $F$ static value of 32 ($p=1.5\times 10^{-14}$). The obtained  hydrogen column density of 
$N_H=1.0(2)\times 10^{22}~\rm{cm^{-2}}$ is more consistent with the previous value. Our fitting indicates a spectral break at
$E_b=5(2)$~keV, and  the  photon index below and above the break are  $\Gamma_1=1.1(2)$ and $\Gamma_2=1.43(3)$, respectively.
The photon index  below the break energy is consistent with the previous  Bepposax observation~\citep{2001A&A...375..397C}  and Chandra observation~\citep{2017ApJ...838..156H}, in which a single power law model is applied to fit the observation data. The photon index above the break ($\Gamma_2\sim 1.43$), on the other hand, is larger than $\sim 1.386(7)$ of the  single power law fitting for NuSTAR  data \citep{2016ApJ...817...93C}.  If the hydrogen column density is fixed to  $N_H=0.9\times 10^{22}~{\rm cm^{-2}}$ of the previous value,
the bk-pl fitting provides $\Gamma_1=0.98(6)$, $E_b=4.3(6)$~keV and $\Gamma_2=1.42(1)$.

Because of enough count rate, we investigate the evolution of the break energy ($E_b$) in the bk-pl model with the spin-phase. We
divide the on-pulsed phase (Figure~\ref{b1509-l}) into six segments,  and we fit each spectrum with a bk-pl model; during the fitting, we fix the hydrogen column density to $N_H=0.9\times 10^{22}~{\rm cm^{-2}}$. The left panel of Figure~\ref{b1509-p} shows the evolution of the  break energy  with the spin phase for PSR~B1509-58. We can see that the break energy is almost constant with $E_{b}=4-8$~keV,
although a small variation may exist.

 We use \verb|cflux| in \verb|Xspec| to estimate the unabsorbed flux
  in 0.3-10~keV bands for each fitting model and  calculate the isotropic luminosity,
  $L_X=4\pi d^2 f_X$, where $f_X$ is the estimated  flux and $d$ is the distance obtained from
  ATNF Pulsar Catalog~\citep{2005AJ....129.1993M}. Figure~\ref{x} compares the radiation
  efficiency ($\equiv \eta_{0.3-10}$) in 0.3-10~keV bands of the MeV pulsar wind those of the $Fermi$-LAT pulsar
  (see  section~\ref{xefficiency} for a detailed discussion).
  To evaluate the radiation efficiency in the wide energy bands,  we also estimated
  the efficiency in 0.3-150~keV using \verb|Xspec|.  We choose the maximum energy
of $150$~keV because there is no HXMT observations for other MeV pulsars and because \verb|Xspec| extrapolates the flux up to 150~keV using the NuSTAR data and the assumed spectral shape.
The estimated luminosity  and its efficiency ($\equiv \eta_X$) in 0.3-150~keV bands  are summarized in Table~2.  For PSR~B1509-58, we obtain the
efficiency of $\eta_X\sim 0.03-0.04$ in 0.3-150~keV bands.

\begin{deluxetable*}{cccccccc}
  \tablecolumns{8}
  \tabletypesize{\footnotesize}
  \tablecaption{Spectral parameters for s-pl  fitting  or bk-pl fitting.}
  \tablehead{
    \colhead{PSR}  &
    \colhead{$N_H$}&
    \colhead{$\Gamma_1^{(a)}$} &
    \colhead{$E^{(b)}_b$} &
    \colhead{$\Gamma^{(c)}_2$} &
    \colhead{$\chi^2$/D.O.F.}  &
    \colhead{$L_X^{(d)}$} &
    \colhead{$\eta^{(e)}_X$}  \\
    \colhead{} &
    \colhead{($10^{22}~{\rm cm^{-2}}$)} &
    \colhead{}&
    \colhead{(keV)}&
    \colhead{}&
    \colhead{}&
    \colhead{$(10^{35}~{\rm erg~s^{-1}})$} &
    \colhead{$10^{-2}$}
  }
  \startdata
  B1509-58 & 1.32(5)  & 1.41(1) & & & 2023.99/2023    & 6.1(2)  & 3.8(1)  \\
  & 1.0(2)  & 1.1(2)  & 5(2)  & 1.43(3) & 1961.24/2021 & 5.7(4) & 3.3(2) \\
  \hline
  J1617-5055 & 4.0(7)  & 1.52(5) & & & 306.63/327 & 0.79(7) & 0.50(5)  \\
  \hline
  J1811-1925  & 4.0(8) & 1.40(8) && & 250.11/287 & 0.8(1)& 1.3(2)    \\
  \hline
  J1813-1749  & 13(4)  &1.8(3)  & &  & 110.29/128   & 0.5(2) & 0.08(4)     \\
  \hline
  J1838-0655 &6.9(2)  & 1.40(2)&&  &  1201.15/1133  & 3.8(2) & 6.9(4) \\
  & 6.0(3) & 1.04(9) & 8.1(9) & 1.49(4) & 1151.75/1131 &3.4(2) & 6.1(4) \\
  \hline
  J1846-0258 & 6(2)  & 1.4(1) & & & 569.09/560    &  1.3(4)& 1.6(4)  \\
  & 5(2) & 1.0(3) &  10(5) & 1.5(1)  & 565.53/558 & 0.7(2) & 0.9(2) \\
  \hline
  J1849-0001 & 5.1(3)  & 1.48(5) & & & 335.95/317    & 2.5(2) & 2.5(2)$^{(f)}$   \\
  & 4.0(9) & 0.8(4) &  4.9(8) & 1.50(6)  & 319.16/315  & 2.3(3) & 2.3(2) \\
  \hline
  J1930+1852 & 2.9(6) & 1.53(8)&  & &  443.32/412    & 0.9(2) & 0.8(2) \\
  & $1.95^{g}$ &  0.8(4)  &  5(2)  &  1.5(1)  & 400.49/409 & 0.8(2) & 0.7(2)
  \enddata
  \tablenotetext{\rm a}{Photon index for the s-pl fitting  or index below the break energy for the bk-pl fitting.}
  \tablenotetext{\rm b}{Break energy for the bk-pl fitting.}
  \tablenotetext{\rm c}{Photon index above break energy for the bk-pl fitting.}
  \tablenotetext{\rm d}{Luminosity in 0.3-150~keV energy bands, $L_X=4\pi d^2 F_{0.3-150~{\rm keV}}$.}
  \tablenotetext{\rm e}{Efficiency of the emission in  0.3-150~keV energy bands.}
  \tablenotetext{\rm f}{Distance is assumed to be $d=7$~kpc~\citep{2011ApJ...729L..16G}.}
  \tablenotetext{\rm g}{Fixed to the value reported in \cite{2010ApJ...710..309T}.}
\end{deluxetable*}
\subsection{PSR~J1657-5055}
\label{psrj1657}
XMM-Newton observed the target with the timing-mode of the PN camera. The spin signal $P_{s}\sim 69.4$~ms is confirmed in the $Z_1^2$-periodogram. Because of the one-dimensional imaging capability of the data taken by the timing mode, we select the pixels so that the extracted events maximize  the  power of the periodic signal. There is one archival data of the NuSTAR observation, but no  observation has been carried out by the NICER.  The right panel of Figure~\ref{b1509} shows the extracted spectrum of the pulsed component. The spectrum is well fitted by a s-pl model with $N_H=4.0(7)\times 10^{22}~{\rm cm^{-2}}$ and $\Gamma_1\sim 1.52(5)$. We find that the best fitting hydrogen column density is slightly smaller than $N_H=6.5(7)\times 10^{22}~{\rm cm^{-2}}$
reported in \cite{2021ApJ...923..249H}, in which only  NuSTAR data is used. 
The photon index $\Gamma_1=1.52(5)$, on the other hand,
is consistent with $\sim 1.51(2)$ of the previous study. 
As the right panel of Figure~\ref{b1509} indicates,
the quality of the data is insufficient to constrain the parameters of the bk-pl model.

The spectrum of the NuSTAR data covers the energy range up to $\sim 70$~keV, and no spectrum cut-off has been observed in 10-70~keV bands. By assuming that  the spectrum above 70~keV extends with
a single power law, we use $\verb|Xspec|$ to extrapolate the  flux up to the energy  150~keV and obtain the efficiency of $\eta_{X}=0.0050(5)$ in 0.3-150~keV bands.

In the spectrum taken by NuSTAR in Figure~\ref{b1509}, one may notice an absorption feature at $\sim 50$~keV, where the data points of both NuSTAR's FPMA and FPMB detectors drop from the average of the residual. We will  also observe  a similar   feature in the spectrum of PSR J1849-0001 with  a higher  significance level (section~3.7). In this study, we apply the Bayesian method to evaluate the  absorption-line feature and  utilize a python tool, \verb|Bayesian  X-ray Analysis| \citep[BXA,][]{2014A&A...564A.125B}. By grouping the NuSTAR's spectra to ensure at least one count per spectral bin, a s-pl with Gaussian absorption model is compared with a pure s-pl model. For PSR J1617-5055, we find that
the inclusion of the absorption feature does not improve the fitting results and could  not give  an evidence of the absorption feature.

\subsection{PSR~J1811-1925}
For PSR J1811-1925, there are one archival data taken by the  XMM-Newton PN camera under  the small window mode and one archival data by NuSTAR. We also use the NICER observations carried out in 2022 March to September.
We fit the pulsed spectrum (Figure~\ref{j1838} and Table~2)
with a s-pl model and obtains $N_H=4.0(8)\times 10^{22}~{\rm cm^{-2}}$ and $\Gamma_1=1.40(8)$, which is consistent with the previous study
using Chandra and NuSTAR data~\citep{2020ApJ...889...23M}. Similar to PSR~J1657-5055, the
quality of the data is insufficient to constrain the parameters of the bk-pl model.
The isotropic  emission efficiency with the s-pl model is estimated to be  $\eta_X=0.013(2)$  in 0.3-150~keV bands.

\subsection{PSR~J1813-1749}
We use two archival data taken by the XMM-Newton PN camera under  the small window mode and one archival data of NuSTAR. We also analyze the data of the NICER observation in 2019 June-August, and refer the ephemeris provided by \cite{2020MNRAS.498.4396H}. The observed spectrum (Figure~\ref{j1838}) is fitted by a s-pl model with
a large column density of $N_H=13(4)\times 10^{22}~{\rm cm^{-2}}$ and photon index of $\Gamma_1=1.8(3)$. 
Because of the strong absorption, the observed spectrum cannot constrain the parameters of the bk-pl model.  
In order to compare with previous results, we fix the hydrogen column density to $N_H=9.8\times 10^{22}~{\rm cm^{-2}}$. The photon index of the s-pl model is $\Gamma_1=1.5(2)$, which is consistent with $\sim 1.48$ of the INTEGRAL observation reported in \cite{2015MNRAS.449.3827K}.
The efficiency in 0.3-150~keV bands is estimated to be $\eta_X=0.0008(4)$. 

\subsection{PSR~J1838-0655}
Both XMM-Newton and NuSTAR observed the source twice and the observations of the XMM-Newton PN were carried out
under  the small window mode.  
We also analyze the data of the NICER observations carried in 2019, 2020 and 2021. 
The observed spectrum (right panel in Figure~\ref{j1838}) can be fitted by a s-pl model with hydrogen column density of $N_H=6.9(2)\times 10^{22}{\rm cm^{-2}}$ and photon index of $\Gamma_1=1.40(2)$.
We find in Table~2 that the bk-pl model significantly improves the fitting with an $F$ static value of 24 ($p=5\times 10^{-11}$). In the bk-pl model, we obtain a  break energy of $E_{b}=8.1(9)$keV. The photon index below and above the energy break are $\Gamma_1=1.04(9)$ and $\Gamma_2=1.49(4)$, respectively, which
are consistent with the values reported in  \cite{2009MNRAS.400..168L},  who analyze the  Chandra, RXTE and the SUZAKU data. The current hydrogen column density, $N_H\sim 6.0\times 10^{22}~{\rm cm^{-2}}$, is  larger than $N_H\sim 4.5(3.7-5.2)\times 10^{22}~{\rm cm^{-2}}$ obtained with Chandra observation~\citep{2008ApJ...681..515G}, while it is consistent with $6.0(5.6-6.5)\times 10^{22}~{\rm cm^{-2}}$ obtained using SUZAKU data~\citep{2008ApJ...681..515G}. We obtain the efficiency of  $\eta_X\sim 0.06-0.07$ in 0.3-150~keV bands, which is the highest among our targets. 

We perform the phase-resolved spectroscopy using the XMM-Newton and NuSTAR data. 
We extract the spectrum of 0.1 phase interval in the on-pulse phase and fit each spectrum with the bk-pl model by fixing the hydrogen column density to $N_H=6.0\times 10^{22}~{\rm cm^{-2}}$. 
The right panel of Figure~\ref{b1509-p} shows the evolution of the break energy with the spin phase. 
The break energy ($E_b$) in the wing of the pulsed shape ($\sim 0.25$ and $\sim 0.75$ spin phase) cannot be constrained as the figure shows. With the large uncertainty, we cannot discuss the evolution of the break energy with the spin-phase. However, the break energy may be smaller at $\sim 0.6-0.7$ spin-phase of the right tip  on the peak. 

\subsection{PSR J1846-0258}
\label{j1846-main}
PSR~J1846-0258 is known as a high-magnetic field pulsar and it  experienced  a magnetar-like X-ray outburst in 2006 \citep{2008Sci...319.1802G}  and  in 2020 \citep{2021ApJ...911L...6B, 2023ApJ...952..120H}. The X-ray emission after the outburst is fitted  by the thermal emission from the surface plus single power law tail in hard X-ray bands,  and it shows the temporal evolution \citep{2023ApJ...952..120H}. Since we are interested in the emission in a quiescent state, we analyze the data taken before 2020 outburst. We use the data of the XMM-Newton observation (small window mode for PN) and NuSTAR observation carried out in 2017 September.  We also generate the spectrum with the  NICER data taken in 2018, 2019 and 2020, for which we refer to the ephemerides reported in \cite{2023ApJ...952..120H}.

Our fitting parameters  of the s-pl model ($N_H=6(2)\times 10^{22}~{\rm cm^{-2}}$ and $\Gamma_1=1.4(1)$)
are  consistent of previous study,  $N_H=4\times 10^{22}~{\rm cm^{-2}}$ (fixed)  and $\Gamma_1=1.23(9)$ \citep{2021ApJ...908..212G}. The bk-pl model does not improve the fitting and its  break energy, $E_{b}=10(5)$~keV  has  a large error size. Because  of the large uncertainty of the hydrogen column density, we  fix it  to $N_H=4\times 10^{22}~{\rm cm^{-2}}$ determined by Chandra
observation~\citep{2008ApJ...686..508N}. For the bk-pl model, the results of the fitting are 
$\Gamma_1=1.0(2)$, $E_b=8(4)$, and $\Gamma_2=1.5(1)$, and hence
the large uncertainty of the break energy remains. We estimate the emission efficiency as $\eta_X\sim 0.009-0.016$ in 0.3-150~keV bands.

\subsection{PSR J1849-0001}
We use one archival data taken by the XMM-Newton PN camera under  the small-window mode and one data taken by  NuSTAR. For the NICER observation, we construct a local ephemeris for the 2018 data. Although \cite{2019ApJ...877...69B} provide the ephemeris covering  2018 February to September, we extend the ephemeris to cover 2018 February to November. The right panel of Figure~\ref{j1930} shows the extracted spectrum of PSR~J1849-0001.

The s-pl model provides an  acceptable  fitting result  (Table~2), and its parameters are $N_H=5.1(3)\times 10^{22}~{\rm cm^2}$ and $\Gamma_1=1.48(5)$. Although the obtained $N_H$ is smaller than $N_{H}=8.1(2)\times 10^{22}~{\rm cm^2}$ reported in \cite{2024ApJ...960...78K}, who use the same XMM-Newton and NuSTAR data as our study, the photon index is consistent with their $\Gamma_1=1.42(3)$ within the error. The bk-pl model improves the fitting results with
an $F$-static value of 8.2 ($p=0.00031$), and
the photon index ($\Gamma_1=0.8(4)$) below the break energy ($E_b=4.9(8)$~keV)  is consistent with $\sim 1.1$ in the  previous studies~\citep{2011ApJ...729L..16G, 2015MNRAS.449.3827K}, in which XMM-Newton data is fitted by a s-pw model. The obtained $N_H=4.0(9)\times 10^{22}~{\rm cm^{2}}$ is also consistent with $N_H=4.3(6)\times 10^{22}~{\rm cm^{2}}$
of the previous study. Assuming the distance $7$~kpc to the source \citep{2011ApJ...729L..16G},
the efficiency in 0.3-150~keV bands is estimated to be  $\eta_X\sim 0.025$.

\cite{2024ApJ...960...78K} suggest the existence of a peaking of the spectrum at $\sim 60$~keV, which is lower than $\sim 1-5$~MeV of PSRs B1509-58 and J1846-0258. As the left panel in Figure~\ref{j1930} shows, on the other hand, our extracted spectrum may indicate the feature of an  absorption at around 40~keV rather than the peaking of the spectrum. In our study, we group the spectrum of the NuSTAR data  at least 100 counts per spectral bin. The absorption-like feature in the spectrum becomes more obvious if we group the spectrum  with  smaller photon counts per spectral bin.  Similar to the descriptions in section~\ref{psrj1657}, we evaluate the absorption feature with BXA. We fit the NuSTAR spectra with a s-pl plus
Gaussian absorption model, and obtain the energy ($E_{ab}$) and width ($\sigma_{ab}$) of the Gaussian  absorption line
as $E_{ab}=36^{+3}_{-4}~{\rm keV}$ and $\sigma_{ab}=0.6^{+1}_{-0.5}$~keV, respectively. We compare the s-pl with a Gaussian absorption
line model with a pure s-pl model and obtain a  Bayesian factor  of $\sim 2.5$, which is not large enough to claim the strong evidence of the
absorption-line feature; a Bayesian factor of $>$10 is usually  desired to confirm the absorption feature \citep{2014A&A...564A.125B}. If the absorption future were real, the non-thermal emissions from PSR~J1849-0001 would be likely  produced
in the region located near the neutron star surface. This is because  such an absorption feature is possibly associated with a cyclotron absorption
in a highly magnetized and dense region around the neutron star.
Since   the current data  is not  enough to draw any conclusion, it would be desired to investigate the absorption as well as peaking feature in  the spectrum by a  deeper observation.

\subsection{PSR J1930+1852}
XMM-Newton observed the target once  under the large window mode and once under  the full frame mode, for which the timing resolutions are still sufficient enough to carry out the phase-resolved spectroscopy for PSR~J1930+1852. NuSTAR observed the target once, but NICER has not observed the target. As Table~2 shows,
the observed spectrum (Figure~\ref{j1930}) is fitted by a s-pl model with a photon index of $\Gamma_1\sim 1.53(8)$.
For the bk-pl model,  the current data cannot constrain the hydrogen column density, and therefore we fix  the value to  $N_H=1.95\times 10^{22}~{\rm cm^{-2}}$
determined by the Chandra observation~\citep{2010ApJ...710..309T}. We obtain the parameters of $\Gamma_1=0.8(4)$, $E_b=5(2)$~keV and $\Gamma_2=1.5(1)$. With the current large uncertainty, the photon
index below  the break energy is consistent with $\Gamma_1=1.2-1.3$ obtained by the Chandra observation
~\citep{2002ApJ...568L..49L,2010ApJ...710..309T}. The estimated efficiency in 0.3-150~keV bands is  $\eta_X=0.007-0.008$.

\section{Discussion}
\label{discussion}
\subsection{Comparison with $Fermi$-LAT pulsars}
\begin{figure}
  \epsscale{1}
  \centerline{
    \includegraphics[scale=0.6]{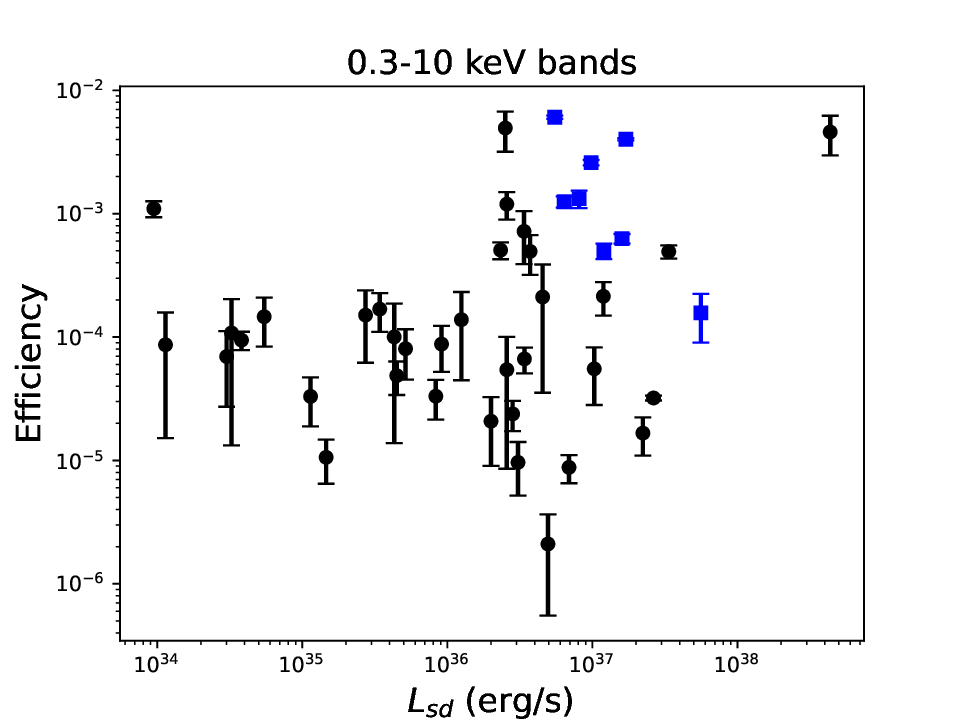}}
  \caption {Efficiency ($\eta_{0.3-10}$) in 0.3-10~keV bands for the MeV pulsars (blue square symbols) and $Fermi$-LAT pulsars (black circle symbols).
    The efficiency of the MeV pulsars  is rescaled to the phase-averaged value by multiplying  the factor of 0.6 to the pulsed flux obtained
  in this study. The $Fermi$ data is taken from \cite{2013ApJS..208...17A}.}
  \label{x}
\end{figure}

\subsubsection{X-ray efficiency}
\label{xefficiency}
Figure~\ref{x} compares the observed efficiency of the isotropic emission  in 0.3-10~keV energy bands $(\eta_{0.3-10}$) of the MeV pulsars with those of the $Fermi$-LAT pulsars, for which we refer the  X-ray flux  of the pulsars
reported in the $Fermi$-LAT second  pulsar catalog~\citep{2013ApJS..208...17A}; in our  study, PSR~B1509-58 is treated as a MeV  pulsar, although it is listed in the LAT pulsar catalog.
Since the $Fermi$-LAT catalog reports the phase-averaged value, we scale the pulsed flux of the MeV pulsars  to the phase-average value
by  multiplying  $0.6$.

From Figure~\ref{x}, we can see the tendency that currently known MeV pulsars
are distributed in a narrower range  of the  spin-down power than the $Fermi$-LAT pulsars, and  no MeV pulsars with $L_{sd}<2\times 10^{36}~{\rm erg~s^{-1}}$ is found. We find that the  MeV pulsars have a larger efficiency than the averaged value ($\sim 10^{-4}$)  of the  $Fermi$-LAT pulsars.  There are six $Fermi$-LAT pulsars  (Crab, PSRs J1119-6127, J1741-2054, J1747-2958, J1801-2451 and J2021+3651), who efficiency is greater than $\eta_{0.3-10}>5\times 10^{-4}$ and has an efficiency similar to the MeV pulsars. We investigate the spectral properties of those $Fermi$-LAT pulsars in the literatures.

Among the $Fermi$-LAT pulsars, the Crab pulsar ($L_{sd}=4.36\times 10^{38}~{\rm erg~s^{-1}}$) has a unique broadband spectrum of the non-thermal emission.
The spectrum has a photon index of $\sim 1.5$ in 0.1-10~keV bands,  $\sim 2$ in 10-100~keV bands and $>2$ in $>100$~keV bands~\citep{2001A&A...378..918K}. The X-ray and soft $\gamma$-ray emissions of the Crab pulsars have been interpreted as a result of the synchrotron emission of the secondary pairs around the light cylinder~\citep{2007ApJ...670..677T, 2015ApJ...811...63H}.
The spectral evolution from X-ray to MeV bands of the Crab pulsar  is  different from those of the MeV pulsars (e.g. PSRs B1509-58. J1838-0655 and J1849-0001), in which the photon index below $E_{b}\sim 5-10$~keV is $\Gamma_1\sim 1$ and $\sim 1.5$ above $E_{b}$ (Table~2). Moreover, $Fermi$-LAT confirmed that the spectra of the two MeV pulsars, PSRs B1509-58 and J1846-0258,
do not extend in the GeV energy band. These differences  may suggest that the emission process and/or emission region of the MeV pulsars is different from that of the Crab pulsar.

For four $Fermi$-LAT pulsars (PSRs J1119-6127, J1741-2054, J1801-2451 and J2021+3651)  with $\eta_{0.3-10}>5\times 10^{-4}$ in 0.3-10~keV bands,
  the non-thermal X-ray spectrum   is described with a photon index of  $\sim 1.6-2.0$~\citep{2008ApJ...680.1417V, 2012ApJ...761...65N,2014ApJ...789...97K, 2014ApJ...790...51M, 2020MNRAS.492.1025C,2021ApJ...917...56B,2023ApJ...954....9W}, which is greater than $\sim 1-1.5$ of MeV pulsars.
  It may be worth to note  that   PSRs J1119-6127 of the  $Fermi$-LAT pulsar and J1846-0258 of the MeV pulsar have been known as high-magnetic field pulsars, and they have experienced the magnetar-like X-ray outbursts~\citep{2008Sci...319.1802G, 2016ApJ...829L..25G,2017ApJ...849L..20A, 2020ApJ...902...96W,2021ApJ...911L...6B}. Their spin-down properties are also resemble to each other; PSR~J1119-6127 has $P_s\sim 407$~ms, $B_s\sim 4.1\times 10^{13}~{\rm G}$ and $L_{sd}\sim 2.3\times 10^{36}~{\rm erg~s^{-1}}$. However, the X-ray emission properties in a quiescent state are different.  The X-ray emission of PSR~J1119-6127 is dominated by thermal component, of  which the   flux in 0.5-8~keV bands is a  several factor more than that of the non-thermal component of $\Gamma_1\sim 2.1$~\citep{2012ApJ...761...65N}.
  For PSR~J1846-0258, on the other hand, the observed X-ray spectrum is described by a non-thermal emission, as we described in \ref{j1846-main}.
 The investigations of PSRs~J1119-6127 and J1846-0258 will suggest that the pulsar spin-down parameter is not key parameters to differentiate between the $Fermi$-LAT pulsars
and the MeV pulsars.

Finally, PSR~J1747-2958 is the  $Fermi$-LAT pulsar having  the highest X-ray efficiency $\eta_{0.3-10}\sim 5\times 10^{-3}$ in Figure~\ref{x}, and its pulsed X-ray emission is reported by \cite{2018ApJ...868L..29L}.
The X-ray emission from the pulsar region taken by the Chandra observation can be  described by a non-thermal emission with a photon index of $1.25-1.55$ \citep{2018ApJ...861....5K,2018ApJ...868L..29L}. 
The relatively hard X-ray spectrum with a high efficiency of PSR~J1747-2958 is similar to the X-ray properties of the  MeV pulsars.  However, because
PSR~J1747-2958 is  associated with bright  pulsar wind nebula (so called Mouse), it cannot exclude the  possibility that a  phase-averaged spectrum reported in the previous studies is
dominated by the nebula emission~\citep{2018ApJ...861....5K}; the phase-resolved spectroscopy is not carried out in the previous studies. We check the archival XMM-Newton PN data and NuSTAR data, but we could not confirm
the pulsed emission in the periodogram. A deeper observation will be required to understand the X-ray emission properties of PSR~J1747-2958.

\begin{figure}
  \epsscale{1}
  \centerline{
    \includegraphics[scale=0.6]{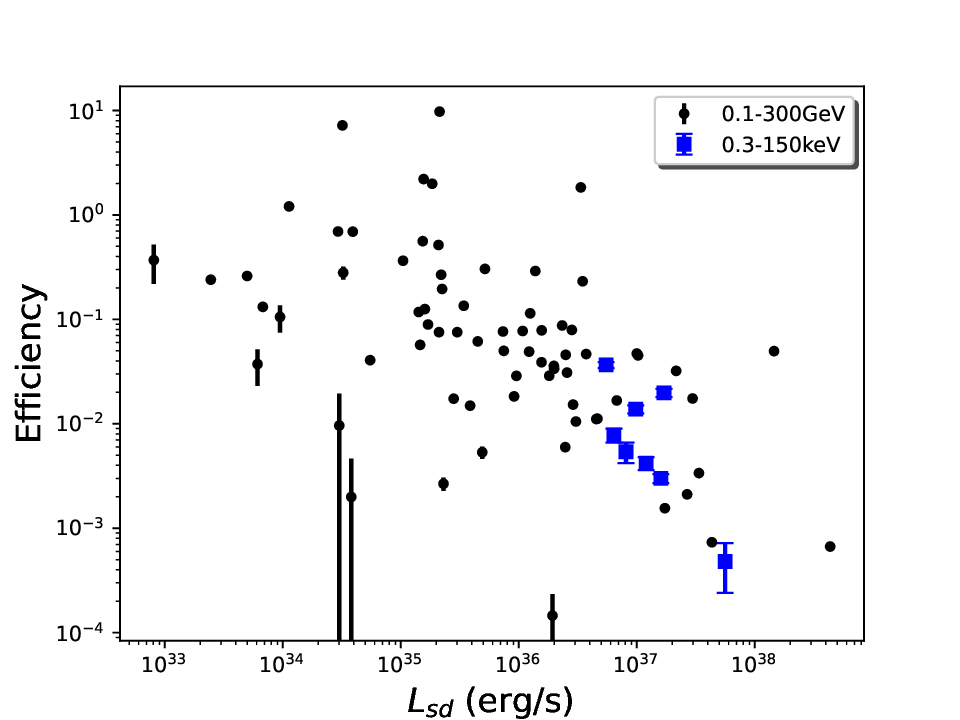}}
  \caption {Comparison between the emission efficiencies of MeV pulsars ($\eta_X$, blue square symbols) in 0.3-150~keV bands and $Fermi$-LAT young pulsars (black circle symbols) in 0.1-300~GeV bands.  The efficiency of the MeV pulsars  is rescaled to the phase-averaged value by multiplying  the factor of 0.6 to the pulsed flux obtained in this study. The $Fermi$ data is taken from \cite{2013ApJS..208...17A}.}
  \label{g}
  \end{figure}

\subsubsection{Efficiency of total non-thermal  pulsed emission}

Multi-wavelength observations have revealed that the non-thermal pulsed  emissions from the MeV pulsars and $Fermi$-LAT pulsars have in general  the most radiation efficiency in $1-10$~MeV and 1-10~GeV bands, respectively. For PSR~B1509-58 of the MeV pulsar, the observed flux in 100~keV-10~MeV bands \citep{1999A&A...351..119K}
indicates that the isotropic efficiency increases by about factor of two to three  from the value ($\eta_X$) reported
in Table~2. By assuming that other MeV pulsars have a  similar spectral behavior to PSR~B1509-58, the typical
efficiency of the MeV pulsar would be estimated to be  $\eta_{MeV}\sim 2-3\times \eta_X\sim 0.01-0.1$;
the isotropic efficiency of PSR~J1813-1749 could be still smaller than 0.01.

Since the efficiency ($\eta_X$) in the 0.3-150~keV bands in Table~2 represents, in order of magnitude, the total efficiency of the MeV pulsars, we compare the efficiencies of the eight MeV pulsars with the efficiencies in 0.1-300~GeV bands of the $Fermi$-LAT pulsars in Figure~\ref{g}.
We plot the efficiency mainly referred to the bk-pl model. 
If the spectra cannot be fitted by the bk-pl, we then derive the value from the s-pl model instead. We note that few $Fermi$-LAT pulsars appears with an efficiency greater than unity. This is likely  due to over-estimated distance and/or the assumption of the isotropic radiation~\citep{2013ApJS..208...17A}. 

Based on Figure~\ref{g}, we may conclude that  the efficiency of the total non-thermal pulsed emission of the MeV pulsars is similar to those of the $Fermi$-LAT pulsars having
  similar spin-down powers. 
Even though the total efficiency of the MeV pulsar could increase by a factor of two to three from the plotted values, it is still consistent with that of the $Fermi$-LAT pulsars. This similarity of the efficiency may suggest that the emission of the MeV pulsars is either (i) a  primary emission but the maximum energy of the electrons/positrons is smaller than that of $Fermi$-LAT pulsars or (ii) a  secondary emission from the electron/positron pairs, into which most of  the  primary GeV emission are converted.

\subsection{Emission process}
In the last section, we estimate that if the emission of the MeV pulsars are isotropic, the typical efficiency is $\eta_{MeV}\sim 10^{-2}-10^{-1}$.
The actual radiation geometry will depend  on various factors such as the emission process, inclination angle and the size of the emission region, and it is not trivial to determine the true efficiency from the
observed flux.  Since  the observed pulsed width covers  $\sim$ 60~\% of the one rotation, the solid angle of the radiation may not be much smaller than the isotropic value of $4\pi$. Hence we assume the efficiency
$\eta_{MeV}\sim 10^{-2}-10^{-3}$ as a more conservative value.
The actual efficiency  could take an  even smaller value, if  the line of sight,
the axis of the spin  and axis of the radiation beam  were  almost aligned  in the same direction. In this paper, however, we do not consider such a peculiar situation.

In the followings, we discuss either the possibility of the curvature radiation process or the synchrotron radiation process. The inverse-Compton process (IC process) can also produce the high-energy photons in the pulsar magnetosphere. For the young pulsars, however, it has been predicted that the energy of the scattered photons are in the TeV energy bands, which can explain the recent observation of
the Vela pulsar~\citep{2023NatAs.tmp..228H}. \cite{2014ApJ...787..167N}  study non-thermal X-ray emission from the millisecond pulsars whose X-ray pulse is in phase with the gamma-ray pulse.
They suggest that  the IC process of the  accelerated electrons (or positrons) off the radio wave
produces the non-thermal X-ray emission. However, this model predicts an accompany  of the GeV emission and is incompatible with the non-detection of
the GeV emission of  the MeV pulsars. We, therefore, will not discuss the IC process as the origin of the MeV pulsar's emission in the following discussion.

\subsubsection{Curvature radiation process}

It is widely accepted that the GeV emission is produced in the region located  near or beyond the light cylinder, because the GeV photons produced near the stellar surface cannot avoid the pair-creation process with the magnetic field.
For canonical ``gap'' models of  the pulsar magnetosphere~(\citealp[eg.][]{1986ApJ...300..500C,1981IAUS...95...69A}), the GeV emission is produced by
the curvature radiation process, due to the bending of the magnetic field lines,  of the primary electrons and/or positrons accelerated by the electric field along the magnetic field line. For this model, the number  density in the acceleration
region cannot be much bigger than the Goldreich-Julian value,  $n_{GJ}$. The characteristic  energy and radiation power of the curvature radiation process from single electron (or positron) are provided by $E_{curv}=3hc\gamma^3/(4\pi R_c)$ and $P_{curv}=2e^2c\gamma^4/3R_c^2$, where $\gamma$ is the Lorentz factor of the particles, $h$ is the Planck constant  and $R_c$ is the curvature radius of the magnetic field line.

One can show that when the dynamical timescale $\tau_{dyn}\sim R_c/c$ is equal to the cooling timescale of the curvature radiation process,
$\tau_{curv}\sim 3m_ecR^2_c/(2e^2\gamma^3)$, the characteristic energy of the radiation is $E_{curv}\sim 9m_ec^2/(4\alpha)\sim 158$~MeV, where $m_e$ the mass of an electron and  $\alpha$ is the fine structure constant. Hence, if the curvature radiation is the origin of the non-thermal emission in X-ray to soft $\gamma$-ray bands of the MeV pulsars, it will be taken place in a slow cooling regime.

In the slow cooling regime, the fraction of the energy of the particles that is released as  the curvature radiation process during the dynamical timescale will be  $\tau_{curv}/\tau_{dyn}\sim E_{curv}/158~\rm {MeV}$, which is $\sim 0.01$ for $E_{curv}\sim 1$~MeV of the MeV pulsars. If
the emission efficiency of the non-thermal emission of the MeV pulsars is  $\eta_{MeV}\sim 10^{-3}-10^{-2}$, it is necessary that  $10\sim 100$~\% of the spin-down energy is converted into the particle energy within or near the light cylinder. Moreover, we can also show that the required multiplicity ($\kappa$) and Lorentz factor $(\gamma)$ to explain the observations ($E_{curv}\sim 1$~MeV and $\eta_{MeV}\sim 10^{-3}-10^{-2}$)  are $\kappa\sim 10^{2-5}$ and $\gamma\sim 10^{5-6}$, respectively. Such a  high-multiplicity indicates that the emission originates from so-called  ``secondary'' electrons/positrons that are
  created by the pair-creation process. To produce the population of the  pairs with  $\gamma\sim 10^{5-6}$ and $\kappa\sim 10^{2-5}$,  it will be required (i)  the pair-creation process of most of an 0.1-1~TeV gamma-rays that are produced with an efficiency of $0.1-1$ or (ii)
  an acceleration of the secondary pairs if they are created by the GeV gamma-rays.

\subsubsection{Synchrotron radiation}
\label{syncrad}
\begin{figure}
  \epsscale{1}
  \centerline{
    \includegraphics[scale=0.6]{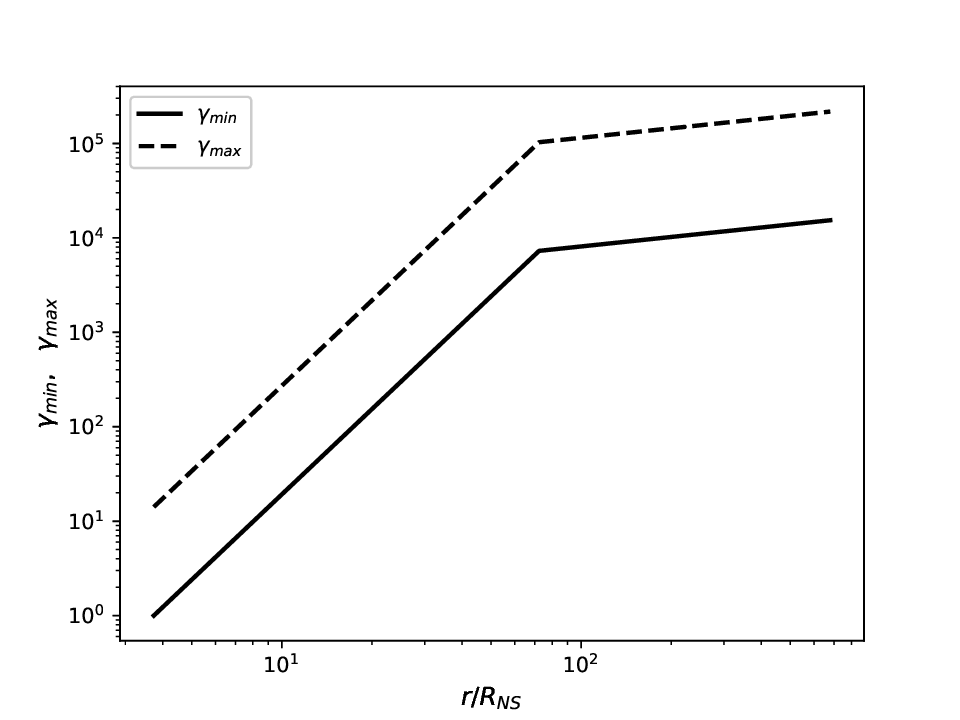}}
  \caption {Parameters of the synchrotron emission model for the MeV pulsars. The solid line and dashed line show  the minimum and the maximum  Lorentz factors in the emission region as a function of the assumed radial distance to  the emission region  from the center of the neutron star. A magnetic dipole
    field, $B(r)=B_s(R_s/r)^3$  is assumed to calculate the magnetic field strength at the distance $r$. The results are for the spin-down parameters
  of PSR~B1509-58.}
  \label{gam}
\end{figure}

\begin{figure}
  \epsscale{1}
  \centerline{
    \includegraphics[scale=0.6]{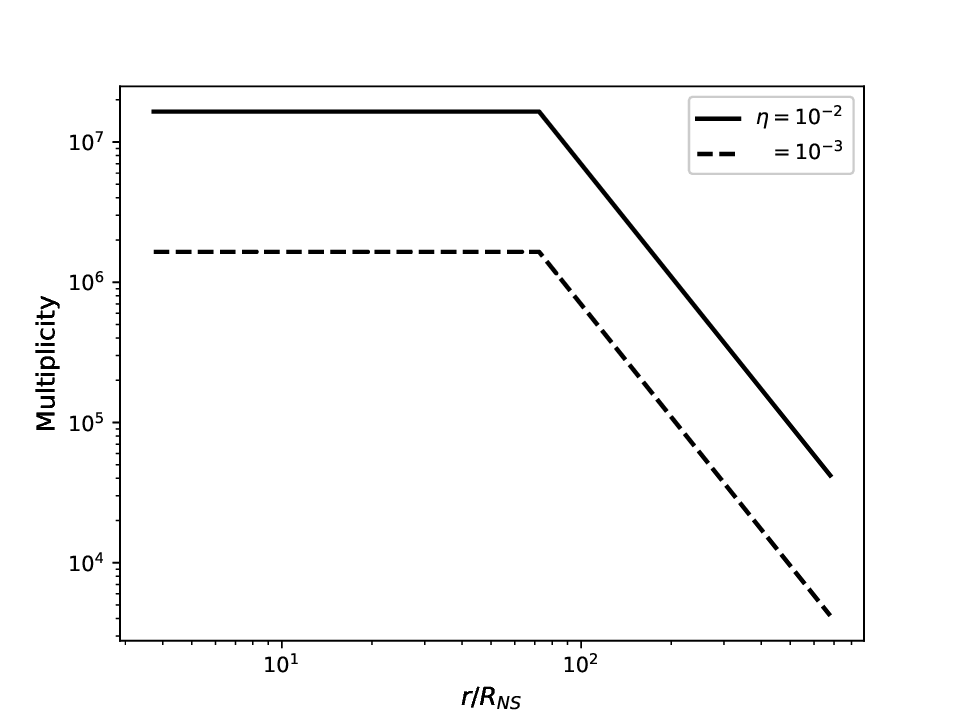}}
  \caption {The required multiplicity of the pairs for the synchrotron emission model of  PSR~B1509-58. The solid line and the dashed line show the
  cases for the radiation efficiency of $\eta_{MeV}=10^{-2}$ and $10^{-3}$, respectively. }
  \label{k}
\end{figure}

The GeV gamma-rays emitted in the magnetosphere could be converted into the electron and positron pairs by the photon-photon pair-creation process or magnetic pair-creation process. The created pairs can lose their energy  via the synchrotron radiation, which is mainly taken place in the fast-cooling regime in the pulsar magnetosphere. In this section, we consider the possibility that the MeV emission originates from the synchrotron radiation from the created pairs. We will not discuss a detailed model of the pair-creation process, while we will  estimate  the required Lorentz factor ($\gamma$), pitch angle ($\theta_p$) and the multiplicity ($\kappa$) of the pairs to explain the spectral properties if the magnetic field strength of the emission region is specified.  Because of insignificant evolution of the pulse shape with the energy bands,
we may make the following assumptions:
\begin{enumerate}
\item Emission in X-ray to soft gamma-ray bands  is produced by the same population of the pairs
  under the synchrotron cooling.
\item The  magnetic field strength and the pitch angle of the pairs do not vary
  in the emission region.
\item The spectral  break  at $E_{b}\sim 5$~keV bands is caused by either (3-i) the cooling timescale, $\tau_{syn}=3m_e^3c^5/(2e^4B^2\gamma_{min} \sin^2\theta_p)$,  equaling to the dynamical timescale of $\tau_{dy}=r/c$ or
  (3-ii) the pairs are already cooled down to the lowest Landau level before running $\tau_{dy}=r/c$.  
\end{enumerate}

For the case of (3-i), if the emission region is located at distance $r$ from the center of the neutron star,
the required Lorentz factor ($\gamma_{min}$) and pitch angle are
\begin{equation}
  \gamma_{min}=\left(\frac{32\pi^2e^2}{27m_ec^4h^2}rE^2_b \right)^{1/3},
~{\rm and~}
  \sin\theta_p= \frac{4\pi m_ec E_b}{3\gamma_{min}^2 ehB},
\label{gmin1}
\end{equation}
respectively, where $B$ is the magnetic field strength in the emission region.  For the case of (3-ii), they are
\begin{equation}
\gamma_{min}=\frac{4\pi m_e c E_b}{3heB},~{\rm and~} \sin\theta_p=1/\gamma_{min},
\label{gmin2}
\end{equation}
respectively. The maximum Lorentz factor  of the pairs, from which the synchrotron cooling starts, are estimated from
\begin{equation}
\gamma_{max}=\left(\frac{4\pi m_ec}{3heB\sin\theta_p}E_{peak}\right)^{1/2}    
\label{gmax}
\end{equation}
where $E_{peak}$ is the peak energy in the spectral energy distribution, and
$E_{peak}=1$~MeV is assumed in this study. Because of the fast cooling regime, the luminosity of the synchrotron emission will be in the order of 
\begin{equation}
  L_{MeV}\sim \gamma_{max}\sin\theta_p m_ec^2 \dot{N}_e,
\end{equation}
where $\dot{N}_e$ is the number of the pairs created per unit time. The ``required'' particle injection rate to explain the observed luminosity is
\begin{equation}
  \dot{N}_e=\frac{\eta_{MeV}L_{sd}}{\gamma_{max}\sin\theta_p m_ec^2}. 
\label{ne}
\end{equation}

In this section, we calculate the required multiplicity, $\kappa$, from the ratio of $\dot{N}_e$ to a particle injection rate from the polar cap  of  the Goldreich-Julian  value, $\dot{N}_{GJ}=2\pi^2B_s R_{NS}^3/(ecP_s^2)$, where $R_{NS}$ is the radius of the neutron star.  By combining the  equations~(\ref{gmin1}) [or (\ref{gmin2})], (\ref{gmax}) and (\ref{ne}) and using  the observed parameters of $E_{b}\sim 5$~keV, $E_{peak}\sim 1$~MeV and $\eta_{MeV}\sim 0.01-0.001$, we can estimate
$\gamma_{min}$, $\sin\theta_p$, $\gamma_{max}$ and $\dot{N}_e$ as a function of
the magnetic field strength. We note that  the required multiplicity is roughly proportional to $(E_{peak}/E_b)^{1/2}$ if we apply the different values for  $E_{peak}$ and/or $E_b$ [see equation (\ref{kappa})].

Figure~\ref{gam} represents the expected $\gamma_{min}$ (solid line) and $\gamma_{max}$ (dashed line) for PSR~B1509-58 as a function of the radial distance from the center of the neutron star, where we assume the dipole magnetic field of $B(r)=B_s(R_{NS}/r)^3$. Figure~\ref{k} shows the required multiplicity by assuming  the emission efficiency of $\eta_{MeV}=10^{-2}$ (solid line) and $10^{-3}$ (dashed line), respectively; we do not plot for $r/R_{NS}<4$, since $\gamma_{min}$ calculated with the dipole magnetic field  for PSR~B1509-58 becomes smaller than unity.
In both figures, we can see a change in the  slopes of the lines at about $r/R_{NS}=80$. This is because that
the parameters  $(\gamma_{min},~\sin\theta_p)$
are determined by  the equation~(\ref{gmin1}) for  $r/R_{NS}>80$,
while it is determined by the equation~(\ref{gmin2}) for  $r/R_{NS}<80$.
As Figure~\ref{k} shows, we can find  that
the required multiplicity in $r/R_{NS}<80$ for  the case~(3-ii)
is independent on the magnetic field strength of the emission region,  and it is expressed by 
\begin{equation}
  \kappa=\eta \frac{L_{sd}}{\dot{N}_{GJ}m_ec^2}\left(\frac{E_{peak}}{E_{b}}\right)^{1/2}.
  \label{kappa}
\end{equation}

We can see in Figure~\ref{k} that the required multiplicity  when the emission
region is located at  $r/R_{NS}<80$ is $\kappa>10^6$. Such a large  pair-multiplicity will be realized only by  pair-creation cascade developed in a region near the stellar surface. For example,  \cite{2019ApJ...871...12T} theoretically investigate  the  cascade process of the polar cap accelerator and estimate the maximum possible  multiplicity
as $\kappa_{max}\sim 3\times 10^6$, although the  predicted multiplicity depends on the electric current structure of the acceleration region; for example,  the polar cap model penetrated by a return electric current produces a multiplicity of one or two orders of magnitude smaller than $\kappa_{max}$~\citep{2015ApJ...810..144T}.
As indicated by Figure~\ref{k}, therefore, if the efficiency of PSR~B1509-58 is $\eta> 0.001$ and the emission originates
from $r/R_{NS}<80$, the multiplicity being comparable to or more than the predicted  maximum possible value will be required. We note that the difference in the estimated
multiplicities of the different MeV pulsars is by  several factors,  assuming the efficiency is a constant.

The pair-creation cascade near the stellar surface can be  also initiated if the inward GeV emission produced in the
outer magnetosphere irradiates the strong magnetic field region near the stellar surface~\citep{1997ApJ...487..370Z,2009MNRAS.400..168L}. Such a cascade model can create the pairs with a rate of $\dot{N}_e\sim \eta_{GeV}L_{sd}/2E_{es}$, where $\eta_{GeV}$ is the efficiency of the inward GeV emission and $E_{es}$ is the maximum photon energy that can escape from the cascade region. Assuming $\eta_{GeV}\sim 10^{-2}$ and $E_{es}\sim 5$~MeV, the multiplicity can be of the order of $\kappa\sim 10^{6}$. Uncertainty of the cascade model of the inward emission is the efficiency of the GeV emission and how fraction of the inward emission irradiates near the stellar surface; a detailed modeling will be necessary~\citep{2014MNRAS.445..604W}.
This model predicts that the viewing angle for the MeV pulsar is closer to the spin axis to avoid the GeV emission from the outer magnetosphere. 

For PSR~B1509-58, the energy corresponding to the cyclotron frequency with $B_s\sim 1.5\times 10^{13}$~G is $heB_s/(2\pi m_ec)\sim 160$~keV. The current synchrotron emission model, therefore, expects that the emission region is located at  $r/R_{NS}>4$, as indicated in Figure~\ref{gam}.
  It may be difficult to realize the multiplicity $\kappa> 10^6$ in the pair-creation cascade developed at $r/R_{NS}>4$ due to the weaker magnetic
  field strength~\citep{2019ApJ...871...12T}. Hence if the observed emission originates from $r/R_{NS}>4$ with a multiplicity $\kappa>10^6$ ($r/R_{NS}<80$), the emitting pairs must be created in the region located near the stellar surface.  Moreover, the pairs created near the surface immediately lose the perpendicular momentum by the synchrotron radiation. Hence, an additional  energy injection processed at the vicinity of the emission region  is required
to gain the perpendicular momentum. For example, \cite{2015ApJ...811...63H} argue that the pairs streaming from the polar cap region gains the
pitch angles through the resonant cyclotron absorption of radio waves.

As Figures~~\ref{gam} and~\ref{k} show, the required  particle multiplicity decreases as the emission region moves farther away from the  neutron star,
while the required Lorentz factor increases. For example, if the emission is produced near the light cylinder, the maximum  Lorentz factor of the pairs is of
the order of $\gamma_{max}\sim 10^5$. To explain the MeV pulsar's emission with the synchrotron emission of the pairs created outer magnetosphere,
$E_{GeV}\sim 100$~GeV of the photon energy with an
emission efficiency $10^{-3}-10^{-2}$ will be required.
Such an emission component has not been  observationally and theoretically predicted.
Hence,  we still need to invoke   an energy injection to the pairs that happen in the outer
magnetosphere.

As we have seen that if the emission of the MeV pulsars originates from the synchrotron radiation from the region inside the light cylinder, the primary GeV emission, which is produced in either the polar cap region or outer magnetosphere, is necessary to create the electron/positron pairs. Within this framework of the scenario, therefore,
it is not  unexpected that the MeV pulsars also produce a GeV emission in the outer magnetosphere, which does not direct
toward the Earth. Considering that the GeV emission in the outer magnetosphere   will be more concentrated to the equatorial plane~\citep{2011ApJ...727..123W, 2018ApJ...857...44K},  we speculate that
the MeV pulsars have in general a smaller viewing angle than that of the
$Fermi$-LAT pulsars, as suggested by \cite{2014MNRAS.445..604W}.

We note that the pulse profile of the non-thermal X-ray emission from the $Fermi$-LAT pulsars (e.g. the Vela pulsar and Crab pulsar) show,
  in general, narrow and multiple peaks~\citep{2015MNRAS.449.3827K,2023arXiv231203198K}.
  This feature of the pulse profile is different from the signal broad peak of  the MeV pulsars.
  The difference in the pulse profile may also indicate  that the viewing geometry
  and/or  the emission region of the MeV pulsars is different from those of the $Fermi$-LAT pulsars.

\subsubsection{Emission from the current sheet}
It has been discussed that the current sheet in equatorial plane beyond the light cylinder is the origin of the high-energy emission from the pulsars~\citep{2016MNRAS.457.2401C,2023ApJ...943..105H,2023ApJ...959..122C}. In this section, therefore,
we estimate the required multiplicity ($\kappa$) of the MeV pulsars with a  current sheet model for the MeV pulsars. For the particle acceleration due to the magnetic reconnection at the current sheet, the typical energy with the Lorentz factor of the particles may be obtained from the relation
that~\citep{1996A&A...311..172L,2023ApJ...959..122C}
\begin{equation}
  \kappa n_{GJ}\times \gamma'_p m_ec^2\sim \Gamma_{bulk}\frac{B_{lc}^2}{8\pi},
  \label{current1}
\end{equation}
where $\Gamma_{bulk}$ is the Lorentz factor of the the bulk flow of the electron/positron pairs,
$\gamma'_p$ is the Lorentz factor of the pairs in the flow rest frame  and $B_{lc}$ is the magnetic field strength at the light cylinder.
The characteristic energy of the synchrotron radiation from the current sheet is written as
\begin{equation}
  E_{peak}\sim \Gamma_{bulk}\times \left(\frac{3}{2}\right)^{3/2}\frac{\gamma'^2_p h eB_{lc}}{2\pi m_ec}.
  \label{current2}
\end{equation}
Using the  equations~(\ref{current1}) and~(\ref{current2}), the multiplicity is estimated from 
\begin{eqnarray}
  \kappa&\sim &4\times 10^6 \left(\frac{\Gamma_{bulk}}{10}\right)^{3/2} \nonumber \\
 && \left(\frac{E_{peak}}{1~{\rm MeV}}\right)^{-1/2}\left(\frac{P_s}{152~{\rm ms}}\right)
    \left(\frac{B_{lc}}{4\cdot 10^4~{\rm~G}}\right)^{3/2}, 
    \label{ksheet}
\end{eqnarray}
where we scale the value using the typical bulk Lorentz factor of $\Gamma_{bulk}=10$~\citep{1996A&A...311..172L} and the pulsar parameters of PSR B1509-58.
For MeV pulsars, the multiplicity with $\kappa\sim 10^6$ will be required if the emission originates from the current sheet.

The rate of the energy injection to the particles from  the current sheet  may be written as~\citep{1996A&A...311..172L}
\begin{equation}
  L\sim \left[ c\left(\frac{\gamma'_p  m_ec^2}{aeB}\right)^{3/2}\frac{B_{lc}^2}{2\pi}\right]\times 2\pi \Gamma_{bulk}R_{lc}x,
\label{current}
\end{equation}
where $x$ is the radial extent of the energy dissipation region and   $a$ represents the sheet width that may be described by
$a\sim (3/\sqrt{r_e})(c/\omega_B)^{3/2}$, where $r_e=e^2/(m_ec^2)$ and $\omega_B=eB/m_ec$ is the classical electron radius
and gyrofrequency, respectively. By   anticipating
that the energy injection  rate of the equation~(\ref{current}) corresponds to the radiation luminosity, $L=\eta_{MeV} L_{sd}$, the required radial size of the current
sheet is estimated to be 
\begin{eqnarray}
  \frac{x}{R_{lc}}&\sim& 0.5\left(\frac{\Gamma_{bulk}}{10}\right)^{-1/4}
  \left(\frac{P_s}{152~{\rm s}}\right)^{2}\left(\frac{B_{lc}}{4\cdot 10^4~{\rm~G}}\right)^{-2} \nonumber \\
  && \left(\frac{E_{peak}}{1~{\rm MeV}}\right)^{3/4}\left(\frac{\eta_{MeV} L_{sd}}{1.7\cdot 10^{34}~{\rm erg~s^{-1}}}\right), 
\end{eqnarray}
which will provide  a reasonable size of $x\sim R_{lc}$. 

For the current sheet model, the difference between $Fermi$-LAT and MeV pulsars will be attributed to the difference in the multiplicities of the pairs injected into the current sheet. The equation~(\ref{ksheet}) implies that the multiplicity for  the $Fermi$-LAT pulsar is of the order of $\kappa\sim 10^{5}$ with
a photon energy of $E_{peak}\sim 1$~GeV, and it is more than  one order of magnitude smaller than that of the MeV pulsar. 
A young pulsar may be observed as a MeV pulsar if the multiplicity of the current sheet is $\kappa\sim 10^6$, while it becomes
a GeV pulsar if $\kappa\sim 10^5$.  One possibility of such a difference
in  the injected multiplicity is the difference in the inclination angle of the magnetic axis from the spin axis.
As we described in section~4.2.2, the large multiplicity of $\kappa\sim 10^6$  can be realized in the cascade process in the polar cap region. Although a detailed study is required, (i) the number of  the created pairs in the polar cap region will depend on the magnetic field lines
and (ii) how  fraction of the created pairs injected into the current sheet depends on the magnetic inclination angle. Intuitively, since a pulsar  with  a larger  magnetic inclination has a smaller  angle between the magnetic axis and equatorial plane, it may inject more pairs to  the equatorial current sheet.  The typical
inclination angle of the  MeV pulsar, therefore,  may be greater than that of the $Fermi$-LAT pulsars. It will be worth to carry out the study of the polar cap cascade process with the effect of the magnetic inclination angle to investigate this possibility.  

\section{Summary}
\label{summary}
We reported on the pulsed X-ray emission properties  of the  MeV pulsars, as measured by  XMM-Newton, NICER, NuSTAR
and HXMT. We fit the broadband X-ray spectra with the s-pl model and bk-pl model. For bk-pl model, the spectrum  below
the break energy ($E_b\sim 5-10$~keV) shows a hard spectrum with a photon index of $\Gamma_1$, while it above
the break energy is $\sim 1.5$ (Table~2). In comparison with the X-ray emission of the $Fermi$-LAT
pulsars, the MeV pulsars have in general a harder spectrum and a higher efficiency in 0.3-10~keV energy bands.
By assuming the isotropic emission, we found that total  efficiency of the non-thermal emission of
the MeV pulsars can reach to $\eta_{MeV}\sim 0.01-0.1$, which is similar to the GeV efficiency of the $Fermi$-LAT pulsars having a similar spin-down power. 

We discussed the curvature radiation and synchrotron radiation as a possible emission process of the MeV
pulsars. For the curvature radiation process, it was shown that $\sim10-100$\% of the energy conversion efficiency from the spin-down power to the secondary pairs are required if the emission efficiency is $\eta_{MeV}=0.001-0.01$.
For the synchrotron radiation inside the right cylinder, if the emission region is $r/R_{lc}<80$, the required multiplicity is  $\kappa\sim 10^6$ for $\eta_{MeV}\sim 0.001$, while $\kappa\sim 10^7$ for $\eta_{MeV}\sim 0.01$. 
Such a large pair-multiplicity will be realized only by the pair-creation cascade developed near the stellar surface.
If the synchrotron emission in the outer magnetosphere is the origin of the emission of the
MeV pulsars, an energy injection to  the secondary pairs is necessary. We speculate that if the emission of the MeV pulsars originates from the region inside the magnetosphere, the observer's viewing angle of the MeV pulsars is closer to the spin axis than that of the $Fermi$-LAT pulsars.
 
We also discussed the synchrotron emission of the pairs from the current sheet and estimated the required multiplicity as $\kappa\sim 10^6$ for the MeV pulsars. We argued that based on the current sheet model, the difference between the MeV pulsars and $Fermi$-LAT pulsars is attributed to the difference
in the multiplicity of the pairs injected to the current sheet. We speculated that the MeV pulsars have a larger
magnetic inclination angle, and inject more pairs to the current sheet than  the $Fermi$-LAT pulsars.

\bigskip
 We thank to referee for his/her useful comments and suggestions.
We acknowledge discussion with Dr S. Shibata, K.S. Cheng, H.K. Chang,  and C.P. Hu. We also thanks for Dr. Y.Tou's help with the HXTM data analysis. This work has made use of the data from the Insight-HXMT mission, a project funded by China National Space Administration (CNSA) and the Chinese Academy of Sciences (CAS). J.T. is  supported by the National Key Research and Development Program of China (grant No. 2020YFC2201400) and the National Natural Science Foundation of
China (grant No. U1838102, 12173014). L.C.-C.L. is supported by  NSTC through grants  110-2112-M-006-006-MY3 and 112-2811-M-006-019.
H.H.W. is supported by the Scientific Research Foundation of Hunan Provincial Education Department (21C0343).
S.K. was financially supported by Japan Society for the Promotion of Science Grants-in-Aid for Scientific Research (KAKENHI) Grant Numbers, JP21H01078, JP22H01267, and JP22K03681.

\section*{Data  Availability}
XMM-Newton, NICER and NuSTAR data can obtained from High Energy Astrophysics Science Archive Research  Center
(\url{https://heasarc.gsfc.nasa.gov/})and HXMT data can be obtained from HXMT website (\url{http://hxmten.ihep.ac.cn}).

\bibliography{ref}

\begin{thebibliography}{}
\expandafter\ifx\csname natexlab\endcsname\relax\def\natexlab#1{#1}\fi

\bibitem[{{Abdo} {et~al.}(2010){Abdo}, {Ackermann}, {Ajello}, {Allafort},
  {Asano}, {Baldini}, {Ballet}, {Barbiellini}, {Baring}, {Bastieri}, {Bechtol},
  {Bellazzini}, {Berenji}, {Blandford}, {Bloom}, {Bonamente}, {Borgland},
  {Bregeon}, {Brez}, {Brigida}, {Bruel}, {Buson}, {Caliandro}, {Cameron},
  {Camilo}, {Caraveo}, {Carrigan}, {Casandjian}, {Cecchi}, {{\c{C}}elik},
  {Chekhtman}, {Cheung}, {Chiang}, {Ciprini}, {Claus}, {Cohen-Tanugi},
  {Conrad}, {den Hartog}, {Dermer}, {de Luca}, {de Palma}, {Dormody}, {Silva},
  {Drell}, {Dubois}, {Dumora}, {Farnier}, {Favuzzi}, {Fegan}, {Ferrara},
  {Focke}, {Frailis}, {Fukazawa}, {Funk}, {Fusco}, {Gargano}, {Gehrels},
  {Germani}, {Giglietto}, {Giordano}, {Glanzman}, {Godfrey}, {Gotthelf},
  {Grenier}, {Grondin}, {Grove}, {Guillemot}, {Guiriec}, {Hanabata}, {Harding},
  {Hays}, {Hobbs}, {Horan}, {Hughes}, {J{\'o}hannesson}, {Johnson}, {Johnson},
  {Johnson}, {Johnston}, {Kamae}, {Kanai}, {Kanbach}, {Katagiri}, {Kataoka},
  {Kawai}, {Keith}, {Kerr}, {Kn{\"o}dlseder}, {Kuss}, {Lande}, {Latronico},
  {Lemoine-Goumard}, {Llena Garde}, {Longo}, {Loparco}, {Lott}, {Lovellette},
  {Lubrano}, {Makeev}, {Manchester}, {Marelli}, {Mazziotta}, {McEnery},
  {Michelson}, {Mitthumsiri}, {Mizuno}, {Moiseev}, {Monte}, {Monzani},
  {Morselli}, {Moskalenko}, {Murgia}, {Nakamori}, {Nolan}, {Norris}, {Nuss},
  {Ohno}, {Ohsugi}, {Omodei}, {Orlando}, {Ormes}, {Paneque}, {Panetta},
  {Parent}, {Pelassa}, {Pepe}, {Pesce-Rollins}, {Piron}, {Porter}, {Rain{\`o}},
  {Rando}, {Razzano}, {Rea}, {Reimer}, {Reimer}, {Reposeur}, {Rodriguez},
  {Romani}, {Roth}, {Ryde}, {Sadrozinski}, {Sander}, {Saz Parkinson},
  {Sgr{\`o}}, {Siskind}, {Smith}, {Smith}, {Spandre}, {Spinelli}, {Starck},
  {Strickman}, {Suson}, {Takahashi}, {Takahashi}, {Tanaka}, {Thayer}, {Thayer},
  {Thompson}, {Thorsett}, {Tibaldo}, {Torres}, {Tosti}, {Tramacere},
  {Uchiyama}, {Usher}, {Vasileiou}, {Venter}, {Vilchez}, {Vitale}, {Waite},
  {Wang}, {Weltevrede}, {Winer}, {Wood}, {Yang}, {Ylinen}, {Ziegler}, {Fermi
  LAT Collaboration}, \& {Pulsar Timing Consortium}}]{2010ApJ...714..927A}
{Abdo}, A.~A., {Ackermann}, M., {Ajello}, M., {et~al.} 2010, \apj, 714, 927

\bibitem[{{Abdo} {et~al.}(2013){Abdo}, {Ajello}, {Allafort}, {Baldini},
  {Ballet}, {Barbiellini}, {Baring}, {Bastieri}, {Belfiore}, {Bellazzini},
  {Bhattacharyya}, {Bissaldi}, {Bloom}, {Bonamente}, {Bottacini}, {Brandt},
  {Bregeon}, {Brigida}, {Bruel}, {Buehler}, {Burgay}, {Burnett}, {Busetto},
  {Buson}, {Caliandro}, {Cameron}, {Camilo}, {Caraveo}, {Casandjian}, {Cecchi},
  {{\c{C}}elik}, {Charles}, {Chaty}, {Chaves}, {Chekhtman}, {Chen}, {Chiang},
  {Chiaro}, {Ciprini}, {Claus}, {Cognard}, {Cohen-Tanugi}, {Cominsky},
  {Conrad}, {Cutini}, {D'Ammando}, {de Angelis}, {DeCesar}, {De Luca}, {den
  Hartog}, {de Palma}, {Dermer}, {Desvignes}, {Digel}, {Di Venere}, {Drell},
  {Drlica-Wagner}, {Dubois}, {Dumora}, {Espinoza}, {Falletti}, {Favuzzi},
  {Ferrara}, {Focke}, {Franckowiak}, {Freire}, {Funk}, {Fusco}, {Gargano},
  {Gasparrini}, {Germani}, {Giglietto}, {Giommi}, {Giordano}, {Giroletti},
  {Glanzman}, {Godfrey}, {Gotthelf}, {Grenier}, {Grondin}, {Grove},
  {Guillemot}, {Guiriec}, {Hadasch}, {Hanabata}, {Harding}, {Hayashida},
  {Hays}, {Hessels}, {Hewitt}, {Hill}, {Horan}, {Hou}, {Hughes}, {Jackson},
  {Janssen}, {Jogler}, {J{\'o}hannesson}, {Johnson}, {Johnson}, {Johnson},
  {Johnson}, {Johnston}, {Kamae}, {Kataoka}, {Keith}, {Kerr}, {Kn{\"o}dlseder},
  {Kramer}, {Kuss}, {Lande}, {Larsson}, {Latronico}, {Lemoine-Goumard},
  {Longo}, {Loparco}, {Lovellette}, {Lubrano}, {Lyne}, {Manchester}, {Marelli},
  {Massaro}, {Mayer}, {Mazziotta}, {McEnery}, {McLaughlin}, {Mehault},
  {Michelson}, {Mignani}, {Mitthumsiri}, {Mizuno}, {Moiseev}, {Monzani},
  {Morselli}, {Moskalenko}, {Murgia}, {Nakamori}, {Nemmen}, {Nuss}, {Ohno},
  {Ohsugi}, {Orienti}, {Orlando}, {Ormes}, {Paneque}, {Panetta}, {Parent},
  {Perkins}, {Pesce-Rollins}, {Pierbattista}, {Piron}, {Pivato}, {Pletsch},
  {Porter}, {Possenti}, {Rain{\`o}}, {Rando}, {Ransom}, {Ray}, {Razzano},
  {Rea}, {Reimer}, {Reimer}, {Renault}, {Reposeur}, {Ritz}, {Romani}, {Roth},
  {Rousseau}, {Roy}, {Ruan}, {Sartori}, {Saz Parkinson}, {Scargle}, {Schulz},
  {Sgr{\`o}}, {Shannon}, {Siskind}, {Smith}, {Spandre}, {Spinelli}, {Stappers},
  {Strong}, {Suson}, {Takahashi}, {Thayer}, {Thayer}, {Theureau}, {Thompson},
  {Thorsett}, {Tibaldo}, {Tibolla}, {Tinivella}, {Torres}, {Tosti}, {Troja},
  {Uchiyama}, {Usher}, {Vandenbroucke}, {Vasileiou}, {Venter}, {Vianello},
  {Vitale}, {Wang}, {Weltevrede}, {Winer}, {Wolff}, {Wood}, {Wood}, {Wood}, \&
  {Yang}}]{2013ApJS..208...17A}
{Abdo}, A.~A., {Ajello}, M., {Allafort}, A., {et~al.} 2013, \apjs, 208, 17

\bibitem[{{Aliu} {et~al.}(2008){Aliu}, {Anderhub}, {Antonelli}, {Antoranz},
  {Backes}, {Baixeras}, {Barrio}, {Bartko}, {Bastieri}, {Becker}, {Bednarek},
  {Berger}, {Bernardini}, {Bigongiari}, {Biland}, {Bock}, {Bonnoli}, {Bordas},
  {Bosch-Ramon}, {Bretz}, {Britvitch}, {Camara}, {Carmona}, {Chilingarian},
  {Commichau}, {Contreras}, {Cortina}, {Costado}, {Covino}, {Curtef}, {Dazzi},
  {De Angelis}, {De Cea del Pozo}, {de los Reyes}, {De Lotto}, {De Maria}, {De
  Sabata}, {Delgado Mendez}, {Dominguez}, {Dorner}, {Doro}, {Els{\"a}sser},
  {Errando}, {Fagiolini}, {Ferenc}, {Fernandez}, {Firpo}, {Fonseca}, {Font},
  {Galante}, {Garcia Lopez}, {Garczarczyk}, {Gaug}, {Goebel}, {Hadasch},
  {Hayashida}, {Herrero}, {H{\"o}hne}, {Hose}, {Hsu}, {Huber}, {Jogler},
  {Kranich}, {La Barbera}, {Laille}, {Leonardo}, {Lindfors}, {Lombardi},
  {Longo}, {Lopez}, {Lorenz}, {Majumdar}, {Maneva}, {Mankuzhiyil}, {Mannheim},
  {Maraschi}, {Mariotti}, {Martinez}, {Mazin}, {Meucci}, {Meyer}, {Miranda},
  {Mirzoyan}, {Moles}, {Moralejo}, {Nieto}, {Nilsson}, {Ninkovic}, {Otte},
  {Oya}, {Paoletti}, {Paredes}, {Pasanen}, {Pascoli}, {Pauss}, {Pegna},
  {Perez-Torres}, {Persic}, {Peruzzo}, {Piccioli}, {Prada}, {Prandini},
  {Puchades}, {Raymers}, {Rhode}, {Rib{\'o}}, {Rico}, {Rissi}, {Robert},
  {R{\"u}gamer}, {Saggion}, {Saito}, {Salvati}, {Sanchez-Conde}, {Sartori},
  {Satalecka}, {Scalzotto}, {Scapin}, {Schweizer}, {Shayduk}, {Shinozaki},
  {Shore}, {Sidro}, {Sierpowska-Bartosik}, {Sillanp{\"a}{\"a}}, {Sobczynska},
  {Spanier}, {Stamerra}, {Stark}, {Takalo}, {Tavecchio}, {Temnikov}, {Tescaro},
  {Teshima}, {Tluczykont}, {Torres}, {Turini}, {Vankov}, {Venturini}, {Vitale},
  {Wagner}, {Wittek}, {Zabalza}, {Zandanel}, {Zanin}, {Zapatero}, {de Jager},
  {de Ona Wilhelmi}, \& {MAGIC Collaboration}}]{2008Sci...322.1221A}
{Aliu}, E., {Anderhub}, H., {Antonelli}, L.~A., {et~al.} 2008, Science, 322,
  1221

\bibitem[{{Archibald} {et~al.}(2017){Archibald}, {Burgay}, {Lyutikov}, {Kaspi},
  {Esposito}, {Israel}, {Kerr}, {Possenti}, {Rea}, {Sarkissian}, {Scholz}, \&
  {Tendulkar}}]{2017ApJ...849L..20A}
{Archibald}, R.~F., {Burgay}, M., {Lyutikov}, M., {et~al.} 2017, \apjl, 849,
  L20

\bibitem[{{Arons}(1981)}]{1981IAUS...95...69A}
{Arons}, J. 1981, in Pulsars: 13 Years of Research on Neutron Stars, ed.
  W.~{Sieber} \& R.~{Wielebinski}, Vol.~95, 69--85

\bibitem[{{Blumer} {et~al.}(2021{\natexlab{a}}){Blumer}, {Safi-Harb},
  {Borghese}, {Mart{\'\i}n}, {McLaughlin}, {Torres}, \&
  {Younes}}]{2021ApJ...917...56B}
{Blumer}, H., {Safi-Harb}, S., {Borghese}, A., {et~al.} 2021{\natexlab{a}},
  \apj, 917, 56

\bibitem[{{Blumer} {et~al.}(2021{\natexlab{b}}){Blumer}, {Safi-Harb},
  {McLaughlin}, \& {Fiore}}]{2021ApJ...911L...6B}
{Blumer}, H., {Safi-Harb}, S., {McLaughlin}, M.~A., \& {Fiore}, W.
  2021{\natexlab{b}}, \apjl, 911, L6

\bibitem[{{Bogdanov} {et~al.}(2019){Bogdanov}, {Ho}, {Enoto}, {Guillot},
  {Harding}, {Jaisawal}, {Malacaria}, {Manthripragada}, {Arzoumanian}, \&
  {Gendreau}}]{2019ApJ...877...69B}
{Bogdanov}, S., {Ho}, W. C.~G., {Enoto}, T., {et~al.} 2019, \apj, 877, 69

\bibitem[{{Buccheri} {et~al.}(1983){Buccheri}, {Bennett}, {Bignami}, {Bloemen},
  {Boriakoff}, {Caraveo}, {Hermsen}, {Kanbach}, {Manchester}, {Masnou},
  {Mayer-Hasselwander}, {{\"O}zel}, {Paul}, {Sacco}, {Scarsi}, \&
  {Strong}}]{1983A&A...128..245B}
{Buccheri}, R., {Bennett}, K., {Bignami}, G.~F., {et~al.} 1983, \aap, 128, 245

\bibitem[{{Buchner} {et~al.}(2014){Buchner}, {Georgakakis}, {Nandra}, {Hsu},
  {Rangel}, {Brightman}, {Merloni}, {Salvato}, {Donley}, \&
  {Kocevski}}]{2014A&A...564A.125B}
{Buchner}, J., {Georgakakis}, A., {Nandra}, K., {et~al.} 2014, \aap, 564, A125

\bibitem[{{Cerutti} {et~al.}(2015){Cerutti}, {Philippov}, {Parfrey}, \&
  {Spitkovsky}}]{2015MNRAS.448..606C}
{Cerutti}, B., {Philippov}, A., {Parfrey}, K., \& {Spitkovsky}, A. 2015,
  \mnras, 448, 606

\bibitem[{{Cerutti} {et~al.}(2016){Cerutti}, {Philippov}, \&
  {Spitkovsky}}]{2016MNRAS.457.2401C}
{Cerutti}, B., {Philippov}, A.~A., \& {Spitkovsky}, A. 2016, \mnras, 457, 2401

\bibitem[{{Chen} {et~al.}(2016){Chen}, {An}, {Kaspi}, {Harrison}, {Madsen}, \&
  {Stern}}]{2016ApJ...817...93C}
{Chen}, G., {An}, H., {Kaspi}, V.~M., {et~al.} 2016, \apj, 817, 93

\bibitem[{{Cheng} {et~al.}(1986){Cheng}, {Ho}, \&
  {Ruderman}}]{1986ApJ...300..500C}
{Cheng}, K.~S., {Ho}, C., \& {Ruderman}, M. 1986, \apj, 300, 500

\bibitem[{{Chernoglazov} {et~al.}(2023){Chernoglazov}, {Hakobyan}, \&
  {Philippov}}]{2023ApJ...959..122C}
{Chernoglazov}, A., {Hakobyan}, H., \& {Philippov}, A. 2023, \apj, 959, 122

\bibitem[{{Coti Zelati} {et~al.}(2020){Coti Zelati}, {Torres}, {Li}, \&
  {Vigan{\`o}}}]{2020MNRAS.492.1025C}
{Coti Zelati}, F., {Torres}, D.~F., {Li}, J., \& {Vigan{\`o}}, D. 2020, \mnras,
  492, 1025

\bibitem[{{Cusumano} {et~al.}(2001){Cusumano}, {Mineo}, {Massaro}, {Nicastro},
  {Trussoni}, {Massaglia}, {Hermsen}, \& {Kuiper}}]{2001A&A...375..397C}
{Cusumano}, G., {Mineo}, T., {Massaro}, E., {et~al.} 2001, \aap, 375, 397

\bibitem[{{Gavriil} {et~al.}(2008){Gavriil}, {Gonzalez}, {Gotthelf}, {Kaspi},
  {Livingstone}, \& {Woods}}]{2008Sci...319.1802G}
{Gavriil}, F.~P., {Gonzalez}, M.~E., {Gotthelf}, E.~V., {et~al.} 2008, Science,
  319, 1802

\bibitem[{{Goldreich} \& {Julian}(1969)}]{1969ApJ...157..869G}
{Goldreich}, P., \& {Julian}, W.~H. 1969, \apj, 157, 869

\bibitem[{{Gotthelf} \& {Halpern}(2008)}]{2008ApJ...681..515G}
{Gotthelf}, E.~V., \& {Halpern}, J.~P. 2008, \apj, 681, 515

\bibitem[{{Gotthelf} {et~al.}(2011){Gotthelf}, {Halpern}, {Terrier}, \&
  {Mattana}}]{2011ApJ...729L..16G}
{Gotthelf}, E.~V., {Halpern}, J.~P., {Terrier}, R., \& {Mattana}, F. 2011,
  \apjl, 729, L16

\bibitem[{{Gotthelf} {et~al.}(2021){Gotthelf}, {Safi-Harb}, {Straal}, \&
  {Gelfand}}]{2021ApJ...908..212G}
{Gotthelf}, E.~V., {Safi-Harb}, S., {Straal}, S.~M., \& {Gelfand}, J.~D. 2021,
  \apj, 908, 212

\bibitem[{{G{\"o}{\u{g}}{\"u}{\c{s}}}
  {et~al.}(2016){G{\"o}{\u{g}}{\"u}{\c{s}}}, {Lin}, {Kaneko}, {Kouveliotou},
  {Watts}, {Chakraborty}, {Alpar}, {Huppenkothen}, {Roberts}, {Younes}, \& {van
  der Horst}}]{2016ApJ...829L..25G}
{G{\"o}{\u{g}}{\"u}{\c{s}}}, E., {Lin}, L., {Kaneko}, Y., {et~al.} 2016, \apjl,
  829, L25

\bibitem[{{H.~E.~S.~S. Collaboration} {et~al.}(2023){H.~E.~S.~S.
  Collaboration}, {Aharonian}, {Ait Benkhali}, {Aschersleben}, {Ashkar},
  {Backes}, {Martins}, {Batzofin}, {Becherini}, {Berge}, {Bernl{\"o}hr}, {Bi},
  {B{\"o}ttcher}, {Boisson}, {Bolmont}, {de Bony de Lavergne}, {Borowska},
  {Bradascio}, {Breuhaus}, {Brose}, {Brun}, {Bruno}, {Bulik},
  {Burger-Scheidlin}, {Bylund}, {Cangemi}, {Caroff}, {Casanova}, {Celic},
  {Cerruti}, {Chand}, {Chandra}, {Chen}, {Chibueze}, {Cotter}, {Mbarubucyeye},
  {Djannati-Ata{\"\i}}, {Dmytriiev}, {Egberts}, {Ernenwein}, {Feijen},
  {Fiasson}, {Fichet de Clairfontaine}, {Fontaine}, {F{\"u}{\ss}ling}, {Funk},
  {Gabici}, {Gallant}, {Ghafourizadeh}, {Giavitto}, {Giunti}, {Glawion},
  {Glicenstein}, {Goswami}, {Grolleron}, {Grondin}, {Haerer}, {Haupt},
  {Hinton}, {Hofmann}, {Holch}, {Holler}, {Horns}, {Huang}, {Jamrozy},
  {Jankowsky}, {Joshi}, {Jung-Richardt}, {Kasai}, {Katarzy{\'n}ski},
  {Kh{\'e}lifi}, {Klepser}, {Klu{\v{z}}niak}, {Komin}, {Kosack}, {Kostunin},
  {Lang}, {Stum}, {Lemi{\`e}re}, {Lemoine-Goumard}, {Lenain}, {Leuschner},
  {Lohse}, {Luashvili}, {Lypova}, {Mackey}, {Malyshev}, {Malyshev}, {Marandon},
  {Marchegiani}, {Marcowith}, {Marinos}, {Mart{\'\i}-Devesa}, {Marx}, {Maurin},
  {Meyer}, {Mitchell}, {Moderski}, {Mohrmann}, {Montanari}, {Moulin}, {Muller},
  {Murach}, {Nakashima}, {de Naurois}, {Niemiec}, {Noel}, {O'Brien}, {Ohm},
  {Olivera-Nieto}, {de Ona Wilhelmi}, {Ostrowski}, {Panny}, {Panter},
  {Parsons}, {Peron}, {Pita}, {Prokhorov}, {Prokoph}, {P{\"u}hlhofer}, {Punch},
  {Quirrenbach}, {Reichherzer}, {Reimer}, {Reimer}, {Renaud}, {Rieger},
  {Rowell}, {Rudak}, {Ruiz-Velasco}, {Sahakian}, {Sailer}, {Salzmann},
  {Sanchez}, {Santangelo}, {Sasaki}, {Sch{\"u}ssler}, {Schwanke}, {Shapopi},
  {Sinha}, {Sol}, {Specovius}, {Spencer}, {Spir-Jacob}, {Stawarz}, {Steenkamp},
  {Steinmassl}, {Steppa}, {Sushch}, {Suzuki}, {Takahashi}, {Tanaka},
  {Tavernier}, {Terrier}, {Thorpe-Morgan}, {Tluczykont}, {Tsirou}, {Tsuji},
  {van Eldik}, {Vecchi}, {Veh}, {Venter}, {Vink}, {Wagner}, {Werner}, {White},
  {Wierzcholska}, {Wun Wong}, {Yassin}, {Zacharias}, {Zargaryan}, {Zdziarski},
  {Zech}, {Zhu}, {Zouari}, {{\.Z}ywucka}, {Zanin}, {Kerr}, {Johnston},
  {Shannon}, \& {Smith}}]{2023NatAs.tmp..228H}
{H.~E.~S.~S. Collaboration}, {Aharonian}, F., {Ait Benkhali}, F., {et~al.}
  2023, Nature Astronomy, doi:10.1038/s41550-023-02151-1

\bibitem[{{Hakobyan} {et~al.}(2023){Hakobyan}, {Philippov}, \&
  {Spitkovsky}}]{2023ApJ...943..105H}
{Hakobyan}, H., {Philippov}, A., \& {Spitkovsky}, A. 2023, \apj, 943, 105

\bibitem[{{Harding} \& {Kalapotharakos}(2015)}]{2015ApJ...811...63H}
{Harding}, A.~K., \& {Kalapotharakos}, C. 2015, \apj, 811, 63

\bibitem[{{Harding} \& {Kalapotharakos}(2017)}]{2017ifs..confE...6H}
{Harding}, A.~K., \& {Kalapotharakos}, C. 2017, in Proceedings of the 7th
  International Fermi Symposium, 6

\bibitem[{{Hare} {et~al.}(2021){Hare}, {Volkov}, {Pavlov}, {Kargaltsev}, \&
  {Johnston}}]{2021ApJ...923..249H}
{Hare}, J., {Volkov}, I., {Pavlov}, G.~G., {Kargaltsev}, O., \& {Johnston}, S.
  2021, \apj, 923, 249

\bibitem[{{Ho} {et~al.}(2020){Ho}, {Guillot}, {Saz Parkinson}, {Limyansky},
  {Ng}, {Bejger}, {Espinoza}, {Haskell}, {Jaisawal}, \&
  {Malacaria}}]{2020MNRAS.498.4396H}
{Ho}, W. C.~G., {Guillot}, S., {Saz Parkinson}, P.~M., {et~al.} 2020, \mnras,
  498, 4396

\bibitem[{{Hsiang} \& {Chang}(2021)}]{2021MNRAS.502..390H}
{Hsiang}, J.-Y., \& {Chang}, H.-K. 2021, \mnras, 502, 390

\bibitem[{{Hu} {et~al.}(2017){Hu}, {Ng}, {Takata}, {Shannon}, \&
  {Johnston}}]{2017ApJ...838..156H}
{Hu}, C.-P., {Ng}, C.~Y., {Takata}, J., {Shannon}, R.~M., \& {Johnston}, S.
  2017, \apj, 838, 156

\bibitem[{{Hu} {et~al.}(2023){Hu}, {Kuiper}, {Harding}, {Younes}, {Blumer},
  {Ho}, {Enoto}, {Espinoza}, \& {Gendreau}}]{2023ApJ...952..120H}
{Hu}, C.-P., {Kuiper}, L., {Harding}, A.~K., {et~al.} 2023, \apj, 952, 120

\bibitem[{{Kalapotharakos} {et~al.}(2018){Kalapotharakos}, {Brambilla},
  {Timokhin}, {Harding}, \& {Kazanas}}]{2018ApJ...857...44K}
{Kalapotharakos}, C., {Brambilla}, G., {Timokhin}, A., {Harding}, A.~K., \&
  {Kazanas}, D. 2018, \apj, 857, 44

\bibitem[{{Kalapotharakos} {et~al.}(2019){Kalapotharakos}, {Harding},
  {Kazanas}, \& {Wadiasingh}}]{2019ApJ...883L...4K}
{Kalapotharakos}, C., {Harding}, A.~K., {Kazanas}, D., \& {Wadiasingh}, Z.
  2019, \apjl, 883, L4

\bibitem[{{Kargaltsev} {et~al.}(2023){Kargaltsev}, {Hare}, \&
  {Lange}}]{2023arXiv231203198K}
{Kargaltsev}, O., {Hare}, J., \& {Lange}, A. 2023, arXiv e-prints,
  arXiv:2312.03198

\bibitem[{{Kargaltsev} {et~al.}(2009){Kargaltsev}, {Pavlov}, \&
  {Wong}}]{2009ApJ...690..891K}
{Kargaltsev}, O., {Pavlov}, G.~G., \& {Wong}, J.~A. 2009, \apj, 690, 891

\bibitem[{{Karpova} {et~al.}(2014){Karpova}, {Danilenko}, {Shibanov},
  {Shternin}, \& {Zyuzin}}]{2014ApJ...789...97K}
{Karpova}, A., {Danilenko}, A., {Shibanov}, Y., {Shternin}, P., \& {Zyuzin}, D.
  2014, \apj, 789, 97

\bibitem[{{Kim} {et~al.}(2024){Kim}, {Park}, {Woo}, {Silverman}, {An}, {Bamba},
  {Mori}, {Reynolds}, \& {Safi-Harb}}]{2024ApJ...960...78K}
{Kim}, C., {Park}, J., {Woo}, J., {et~al.} 2024, \apj, 960, 78

\bibitem[{{Kisaka} \& {Tanaka}(2017)}]{2017ApJ...837...76K}
{Kisaka}, S., \& {Tanaka}, S.~J. 2017, \apj, 837, 76

\bibitem[{{Klingler} {et~al.}(2018){Klingler}, {Kargaltsev}, {Pavlov}, {Ng},
  {Beniamini}, \& {Volkov}}]{2018ApJ...861....5K}
{Klingler}, N., {Kargaltsev}, O., {Pavlov}, G.~G., {et~al.} 2018, \apj, 861, 5

\bibitem[{{Kuiper} \& {Hermsen}(2015)}]{2015MNRAS.449.3827K}
{Kuiper}, L., \& {Hermsen}, W. 2015, \mnras, 449, 3827

\bibitem[{{Kuiper} {et~al.}(2001){Kuiper}, {Hermsen}, {Cusumano}, {Diehl},
  {Sch{\"o}nfelder}, {Strong}, {Bennett}, \& {McConnell}}]{2001A&A...378..918K}
{Kuiper}, L., {Hermsen}, W., {Cusumano}, G., {et~al.} 2001, \aap, 378, 918

\bibitem[{{Kuiper} {et~al.}(2018){Kuiper}, {Hermsen}, \&
  {Dekker}}]{2018MNRAS.475.1238K}
{Kuiper}, L., {Hermsen}, W., \& {Dekker}, A. 2018, \mnras, 475, 1238

\bibitem[{{Kuiper} {et~al.}(1999){Kuiper}, {Hermsen}, {Krijger}, {Bennett},
  {Carrami{\~n}ana}, {Sch{\"o}nfelder}, {Bailes}, \&
  {Manchester}}]{1999A&A...351..119K}
{Kuiper}, L., {Hermsen}, W., {Krijger}, J.~M., {et~al.} 1999, \aap, 351, 119

\bibitem[{{Li} {et~al.}(2018){Li}, {Torres}, {Coti Zelati}, {Papitto}, {Kerr},
  \& {Rea}}]{2018ApJ...868L..29L}
{Li}, J., {Torres}, D.~F., {Coti Zelati}, F., {et~al.} 2018, \apjl, 868, L29

\bibitem[{{Lin} {et~al.}(2009){Lin}, {Takata}, {Hwang}, \&
  {Liang}}]{2009MNRAS.400..168L}
{Lin}, L. C.-C., {Takata}, J., {Hwang}, C.-Y., \& {Liang}, J.-S. 2009, \mnras,
  400, 168

\bibitem[{{Lu} {et~al.}(2007){Lu}, {Wang}, {Gotthelf}, \&
  {Qu}}]{2007ApJ...663..315L}
{Lu}, F., {Wang}, Q.~D., {Gotthelf}, E.~V., \& {Qu}, J. 2007, \apj, 663, 315

\bibitem[{{Lu} {et~al.}(2002){Lu}, {Wang}, {Aschenbach}, {Durouchoux}, \&
  {Song}}]{2002ApJ...568L..49L}
{Lu}, F.~J., {Wang}, Q.~D., {Aschenbach}, B., {Durouchoux}, P., \& {Song},
  L.~M. 2002, \apjl, 568, L49

\bibitem[{{Lyubarskii}(1996)}]{1996A&A...311..172L}
{Lyubarskii}, Y.~E. 1996, \aap, 311, 172

\bibitem[{{Madsen} {et~al.}(2020){Madsen}, {Fryer}, {Grefenstette}, {Lopez},
  {Reynolds}, \& {Zoglauer}}]{2020ApJ...889...23M}
{Madsen}, K.~K., {Fryer}, C.~L., {Grefenstette}, B.~W., {et~al.} 2020, \apj,
  889, 23

\bibitem[{{MAGIC Collaboration} {et~al.}(2020){MAGIC Collaboration}, {Acciari},
  {Ansoldi}, {Antonelli}, {Arbet Engels}, {Asano}, {Baack}, {Babi{\'c}},
  {Baquero}, {Barres de Almeida}, {Barrio}, {Becerra Gonz{\'a}lez}, {Bednarek},
  {Bellizzi}, {Bernardini}, {Bernardos}, {Berti}, {Besenrieder},
  {Bhattacharyya}, {Bigongiari}, {Biland}, {Blanch}, {Bonnoli},
  {Bo{\v{s}}njak}, {Busetto}, {Carosi}, {Ceribella}, {Cerruti}, {Chai},
  {Chilingarian}, {Cikota}, {Colak}, {Colombo}, {Contreras}, {Cortina},
  {Covino}, {D'Amico}, {D'Elia}, {da Vela}, {Dazzi}, {de Angelis}, {de Lotto},
  {Delfino}, {Delgado}, {Delgado Mendez}, {Depaoli}, {di Girolamo}, {di
  Pierro}, {di Venere}, {Do Souto Espi{\~n}eira}, {Dominis Prester}, {Donini},
  {Dorner}, {Doro}, {Elsaesser}, {Fallah Ramazani}, {Fattorini}, {Ferrara},
  {Foffano}, {Fonseca}, {Font}, {Fruck}, {Fukami}, {Garc{\'\i}a L{\'o}pez},
  {Garczarczyk}, {Gasparyan}, {Gaug}, {Giglietto}, {Giordano}, {Gliwny},
  {Godinovi{\'c}}, {Green}, {Green}, {Hadasch}, {Hahn}, {Heckmann}, {Herrera},
  {Hoang}, {Hrupec}, {H{\"u}tten}, {Inada}, {Inoue}, {Ishio}, {Iwamura},
  {Jormanainen}, {Jouvin}, {Kajiwara}, {Karjalainen}, {Kerszberg}, {Kobayashi},
  {Kubo}, {Kushida}, {Lamastra}, {Lelas}, {Leone}, {Lindfors}, {Lombardi},
  {Longo}, {L{\'o}pez-Coto}, {L{\'o}pez-Moya}, {L{\'o}pez-Oramas}, {Loporchio},
  {Machado de Oliveira Fraga}, {Maggio}, {Majumdar}, {Makariev}, {Mallamaci},
  {Maneva}, {Manganaro}, {Mannheim}, {Maraschi}, {Mariotti}, {Mart{\'\i}nez},
  {Mazin}, {Mender}, {Mi{\'c}anovi{\'c}}, {Miceli}, {Miener}, {Minev},
  {Miranda}, {Mirzoyan}, {Molina}, {Moralejo}, {Morcuende}, {Moreno},
  {Moretti}, {Munar-Adrover}, {Neustroev}, {Nigro}, {Nilsson}, {Ninci},
  {Nishijima}, {Noda}, {Nozaki}, {Ohtani}, {Oka}, {Otero-Santos}, {Palatiello},
  {Paneque}, {Paoletti}, {Paredes}, {Pavleti{\'c}}, {Pe{\~n}il}, {Perennes},
  {Persic}, {Prada Moroni}, {Prandini}, {Priyadarshi}, {Puljak}, {Rhode},
  {Rib{\'o}}, {Rico}, {Righi}, {Rugliancich}, {Saha}, {Sahakyan}, {Saito},
  {Sakurai}, {Satalecka}, {Saturni}, {Schleicher}, {Schmidt}, {Schweizer},
  {Sitarek}, {{\v{S}}nidari{\'c}}, {Sobczynska}, {Spolon}, {Stamerra}, {Strom},
  {Strzys}, {Suda}, {Suri{\'c}}, {Takahashi}, {Tavecchio}, {Temnikov},
  {Terzi{\'c}}, {Teshima}, {Torres-Alb{\`a}}, {Tosti}, {Truzzi}, {Tutone}, {van
  Scherpenberg}, {Vanzo}, {Vazquez Acosta}, {Ventura}, {Verguilov}, {Vigorito},
  {Vitale}, {Vovk}, {Will}, {Zari{\'c}}, {Hirotani}, \& {Saz
  Parkinson}}]{2020A&A...643L..14M}
{MAGIC Collaboration}, {Acciari}, V.~A., {Ansoldi}, S., {et~al.} 2020, \aap,
  643, L14

\bibitem[{{Manchester} {et~al.}(2005){Manchester}, {Hobbs}, {Teoh}, \&
  {Hobbs}}]{2005AJ....129.1993M}
{Manchester}, R.~N., {Hobbs}, G.~B., {Teoh}, A., \& {Hobbs}, M. 2005, \aj, 129,
  1993

\bibitem[{{Marelli} {et~al.}(2014){Marelli}, {Belfiore}, {Saz Parkinson},
  {Caraveo}, {De Luca}, {Sarazin}, {Salvetti}, {Sivakoff}, \&
  {Camilo}}]{2014ApJ...790...51M}
{Marelli}, M., {Belfiore}, A., {Saz Parkinson}, P., {et~al.} 2014, \apj, 790,
  51

\bibitem[{{Nasa High Energy Astrophysics Science Archive Research Center
  (Heasarc)}(2014)}]{2014ascl.soft08004N}
{Nasa High Energy Astrophysics Science Archive Research Center (Heasarc)}.
  2014, {HEAsoft: Unified Release of FTOOLS and XANADU}, Astrophysics Source
  Code Library, record ascl:1408.004, ascl:1408.004

\bibitem[{{Ng} {et~al.}(2012){Ng}, {Kaspi}, {Ho}, {Weltevrede}, {Bogdanov},
  {Shannon}, \& {Gonzalez}}]{2012ApJ...761...65N}
{Ng}, C.~Y., {Kaspi}, V.~M., {Ho}, W.~C.~G., {et~al.} 2012, \apj, 761, 65

\bibitem[{{Ng} {et~al.}(2008){Ng}, {Slane}, {Gaensler}, \&
  {Hughes}}]{2008ApJ...686..508N}
{Ng}, C.~Y., {Slane}, P.~O., {Gaensler}, B.~M., \& {Hughes}, J.~P. 2008, \apj,
  686, 508

\bibitem[{{Ng} {et~al.}(2014){Ng}, {Takata}, {Leung}, {Cheng}, \&
  {Philippopoulos}}]{2014ApJ...787..167N}
{Ng}, C.~Y., {Takata}, J., {Leung}, G.~C.~K., {Cheng}, K.~S., \&
  {Philippopoulos}, P. 2014, \apj, 787, 167

\bibitem[{{Romani} {et~al.}(2023){Romani}, {Wong}, {Di Lalla}, {Omodei}, {Xie},
  {Ng}, {Ferrazzoli}, {Di Marco}, {Bucciantini}, {Pilia}, {Slane}, {Weisskopf},
  {Johnston}, {Burgay}, {Wei}, {Yang}, {Zhang}, {Antonelli}, {Bachetti},
  {Baldini}, {Baumgartner}, {Bellazzini}, {Bianchi}, {Bongiorno}, {Bonino},
  {Brez}, {Capitanio}, {Castellano}, {Cavazzuti}, {Chen}, {Cibrario},
  {Ciprini}, {Costa}, {De Rosa}, {Del Monte}, {Di Gesu}, {Donnarumma},
  {Doroshenko}, {Dov{\v{c}}iak}, {Ehlert}, {Enoto}, {Evangelista}, {Fabiani},
  {Garcia}, {Gunji}, {Hayashida}, {Heyl}, {Iwakiri}, {Liodakis}, {Kaaret},
  {Karas}, {Kim}, {Kitaguchi}, {Kolodziejczak}, {Krawczynski}, {La Monaca},
  {Latronico}, {Madejski}, {Maldera}, {Manfreda}, {Marin}, {Marinucci},
  {Marscher}, {Marshall}, {Massaro}, {Matt}, {Middei}, {Mitsuishi}, {Mizuno},
  {Muleri}, {Negro}, {O'Dell}, {Oppedisano}, {Pacciani}, {Papitto}, {Pavlov},
  {Perri}, {Pesce-Rollins}, {Petrucci}, {Possenti}, {Poutanen}, {Puccetti},
  {Ramsey}, {Rankin}, {Ratheesh}, {Roberts}, {Sgr{\'o}}, {Soffitta}, {Spandre},
  {Swartz}, {Tamagawa}, {Tavecchio}, {Taverna}, {Tawara}, {Tennant}, {Thomas},
  {Tombesi}, {Trois}, {Tsygankov}, {Turolla}, {Vink}, {Wu}, \&
  {Zane}}]{2023ApJ...957...23R}
{Romani}, R.~W., {Wong}, J., {Di Lalla}, N., {et~al.} 2023, \apj, 957, 23

\bibitem[{{Smith} {et~al.}(2023){Smith}, {Abdollahi}, {Ajello}, {Bailes},
  {Baldini}, {Ballet}, {Baring}, {Bassa}, {Gonzalez}, {Bellazzini}, {Berretta},
  {Bhattacharyya}, {Bissaldi}, {Bonino}, {Bottacini}, {Bregeon}, {Bruel},
  {Burgay}, {Burnett}, {Cameron}, {Camilo}, {Caputo}, {Caraveo}, {Cavazzuti},
  {Chiaro}, {Ciprini}, {Clark}, {Cognard}, {Corongiu}, {Orestano},
  {Crnogorcevic}, {Cuoco}, {Cutini}, {D'Ammando}, {de Angelis}, {DeCesar}, {De
  Gaetano}, {de Menezes}, {Deneva}, {de Palma}, {Di Lalla}, {Dirirsa}, {Di
  Venere}, {Dom{\'\i}nguez}, {Dumora}, {Fegan}, {Ferrara}, {Fiori},
  {Fleischhack}, {Flynn}, {Franckowiak}, {Freire}, {Fukazawa}, {Fusco},
  {Galanti}, {Gammaldi}, {Gargano}, {Gasparrini}, {Giacchino}, {Giglietto},
  {Giordano}, {Giroletti}, {Green}, {Grenier}, {Guillemot}, {Guiriec},
  {Gustafsson}, {Harding}, {Hays}, {Hewitt}, {Horan}, {Hou}, {Jankowski},
  {Johnson}, {Johnson}, {Johnston}, {Kataoka}, {Keith}, {Kerr}, {Kramer},
  {Kuss}, {Latronico}, {Lee}, {Li}, {Li}, {Limyansky}, {Longo}, {Loparco},
  {Lorusso}, {Lovellette}, {Lower}, {Lubrano}, {Lyne}, {Maan}, {Maldera},
  {Manchester}, {Manfreda}, {Marelli}, {Mart{\'\i}-Devesa}, {Mazziotta},
  {McEnery}, {Mereu}, {Michelson}, {Mickaliger}, {Mitthumsiri}, {Mizuno},
  {Moiseev}, {Monzani}, {Morselli}, {Negro}, {Nemmen}, {Nieder}, {Nuss},
  {Omodei}, {Orienti}, {Orlando}, {Ormes}, {Palatiello}, {Paneque},
  {Panzarini}, {Parthasarathy}, {Persic}, {Pesce-Rollins}, {Pillera}, {Poon},
  {Porter}, {Possenti}, {Principe}, {Rain{\`o}}, {Rando}, {Ransom}, {Ray},
  {Razzano}, {Razzaque}, {Reimer}, {Reimer}, {Renault-Tinacci}, {Romani},
  {S{\'a}nchez-Conde}, {Parkinson}, {Scotton}, {Serini}, {Sgr{\`o}}, {Shannon},
  {Sharma}, {Shen}, {Siskind}, {Spandre}, {Spinelli}, {Stappers}, {Stephens},
  {Suson}, {Tabassum}, {Tajima}, {Tak}, {Theureau}, {Thompson}, {Tibolla},
  {Torres}, {Valverde}, {Venter}, {Wadiasingh}, {Wang}, {Wang}, {Wang},
  {Weltevrede}, {Wood}, {Yan}, {Zaharijas}, {Zhang}, \&
  {Zhu}}]{2023ApJ...958..191S}
{Smith}, D.~A., {Abdollahi}, S., {Ajello}, M., {et~al.} 2023, \apj, 958, 191

\bibitem[{{Takata} \& {Chang}(2007)}]{2007ApJ...670..677T}
{Takata}, J., \& {Chang}, H.~K. 2007, \apj, 670, 677

\bibitem[{{Takata} {et~al.}(2010){Takata}, {Wang}, \&
  {Cheng}}]{2010ApJ...715.1318T}
{Takata}, J., {Wang}, Y., \& {Cheng}, K.~S. 2010, \apj, 715, 1318

\bibitem[{{Temim} {et~al.}(2010){Temim}, {Slane}, {Reynolds}, {Raymond}, \&
  {Borkowski}}]{2010ApJ...710..309T}
{Temim}, T., {Slane}, P., {Reynolds}, S.~P., {Raymond}, J.~C., \& {Borkowski},
  K.~J. 2010, \apj, 710, 309

\bibitem[{{Thompson} {et~al.}(1999){Thompson}, {Bailes}, {Bertsch}, {Cordes},
  {D'Amico}, {Esposito}, {Finley}, {Hartman}, {Hermsen}, {Kanbach}, {Kaspi},
  {Kniffen}, {Kuiper}, {Lin}, {Lyne}, {Manchester}, {Matz},
  {Mayer-Hasselwander}, {Michelson}, {Nolan}, {{\"O}gelman}, {Pohl},
  {Ramanamurthy}, {Sreekumar}, {Reimer}, {Taylor}, \&
  {Ulmer}}]{1999ApJ...516..297T}
{Thompson}, D.~J., {Bailes}, M., {Bertsch}, D.~L., {et~al.} 1999, \apj, 516,
  297

\bibitem[{{Timokhin} \& {Harding}(2015)}]{2015ApJ...810..144T}
{Timokhin}, A.~N., \& {Harding}, A.~K. 2015, \apj, 810, 144

\bibitem[{{Timokhin} \& {Harding}(2019)}]{2019ApJ...871...12T}
---. 2019, \apj, 871, 12

\bibitem[{{Van Etten} {et~al.}(2008){Van Etten}, {Romani}, \&
  {Ng}}]{2008ApJ...680.1417V}
{Van Etten}, A., {Romani}, R.~W., \& {Ng}, C.~Y. 2008, \apj, 680, 1417

\bibitem[{{Wang} {et~al.}(2020){Wang}, {Lin}, {Dai}, {Takata}, {Li}, {Hu}, \&
  {Hou}}]{2020ApJ...902...96W}
{Wang}, H.~H., {Lin}, L.~C.~C., {Dai}, S., {et~al.} 2020, \apj, 902, 96

\bibitem[{{Wang} {et~al.}(2014){Wang}, {Ng}, {Takata}, {Leung}, \&
  {Cheng}}]{2014MNRAS.445..604W}
{Wang}, Y., {Ng}, C.~W., {Takata}, J., {Leung}, G. C.~K., \& {Cheng}, K.~S.
  2014, \mnras, 445, 604

\bibitem[{{Wang} {et~al.}(2013){Wang}, {Takata}, \&
  {Cheng}}]{2013ApJ...764...51W}
{Wang}, Y., {Takata}, J., \& {Cheng}, K.~S. 2013, \apj, 764, 51

\bibitem[{{Watters} \& {Romani}(2011)}]{2011ApJ...727..123W}
{Watters}, K.~P., \& {Romani}, R.~W. 2011, \apj, 727, 123

\bibitem[{{Woo} {et~al.}(2023){Woo}, {An}, {Gelfand}, {Hailey}, {Mori},
  {Mukherjee}, {Safi-Harb}, \& {Temim}}]{2023ApJ...954....9W}
{Woo}, J., {An}, H., {Gelfand}, J.~D., {et~al.} 2023, \apj, 954, 9

\bibitem[{{Zhang} \& {Cheng}(1997)}]{1997ApJ...487..370Z}
{Zhang}, L., \& {Cheng}, K.~S. 1997, \apj, 487, 370

\end{thebibliography}
\appendix
\restartappendixnumbering
\section{Information of observations}
We summarize the information of the XMM-Newton observation in Table~A1 and of the NuSTAR/HXMT observations in Table~A2.  Table~A3 summarizes the NICER observations and the local ephemerides used in this study.

\begin{deluxetable*}{lcclc}
  \tablecolumns{8}
  \tabletypesize{\footnotesize}
  \tablecaption{Journal of the XMM-Newton PN  observation.}
  \label{xmm}
  \tablehead{
    \colhead{PSR}  &
    \colhead{Obs. ID}&
    \colhead{Mode} &
    \colhead{Date} &
    \colhead{Exposure (ks)} 
  }
  \startdata
  B1509-58 & 0128120401  & Small  & 2000 Sep 7  & 25    \\ \hline
  J1617-5055 & 0113050701 & Timing & 2001 Sep 3 & 19 \\ \hline  
  J1811-1925  & 0113050401 & Small  & 2002 Mar 29 & 28    \\\hline
  J1813-1749  & 0552790101   & Small & 2009 Mar 27   & 99        \\
   & 0650310101  & Small  & 2011 Mar 13 & 22 \\
  \hline
  J1838-0655 &  0552950101& Small  &  2008 Oct 16 &  63   \\
  & 0852240101  & Small  & 2019 Oct 18  & 91 \\
   \hline
  J1846-0258 &0795650101& Small  & 2017 Sep  20  & 53     \\
  \hline
  J1849-0001 & 0651930201   &Small & 2011 Mar 23  & 54  \\
  \hline
  J1930+1852 & 0406730101   & Large &  2006 Sep 26  & 54  \\ 
             & 0762980101  & Full Frame & 2016 Mar 27  & 110 
  \enddata
\end{deluxetable*}

\begin{deluxetable}{lclc}
  \tablecolumns{8}
  \tabletypesize{\footnotesize}
  \tablecaption{Journal of the NuSTAR and HXMT  observations.}
  \label{nustar}
  \tablehead{
    \colhead{PSR}  &
    \colhead{Obs. ID}&
    \colhead{Date} &
    \colhead{Exposure (ks)}
  }
  \startdata
       NuSTAR     &        &        &  \\
       B1509-58 & 40024004002 & 2013 Jun 7  & 43    \\
       & 40024002001 & 2013 Jun 8  & 43    \\
       &  40024003001 &2013 Jun 9 & 44    \\
        &  40024001002 &2013 Aug 15 & 34    \\
       \hline
  J1617-5055 & 30301013002 & 2018 Apr 29  & 133 \\\hline
  J1811-1925  & 90201027002  & 2016 Jun 22  & 91   \\\hline
  J1813-1749  & 30364003002   & 2018 Mar 25   &  26        \\
  \hline
  J1838-0655 &  3050101300 & 2019 Oct 17  &  114   \\
  & 30501013004   & 2019 Oct 21   & 53  \\
  \hline
  J1846-0258 &40301004002 & 2017 Sep  17   & 95     \\
  \hline
  J1849-0001 & 40660005001    & 2020 Nov 24  & 57   \\
  \hline
  J1930+1852 & 40101006002   & 2016 Mar 27  &  54    \\
  & 40201012002   & 2016-07-02  & 80     \\ \hline 
  HXMT     &      &        &  \\
  B1509-58 & P0101324001  & 2017 Jul  & 7   
  \enddata
\end{deluxetable}

\begin{deluxetable*}{llllll}
  \tablecolumns{8}
  \tabletypesize{\footnotesize}
  \tablecaption{Ephemerides to fold the NICER data in this study.}
  \label{ephe}
  \tablehead{
    \colhead{PSR}  &
    \colhead{Obs. IDs$^{a}$}&
    \colhead{$f_0$} &
    \colhead{$f_1$} &
    \colhead{$f_2$} &
    \colhead{$T_0$}  \\
    \colhead{} &
    \colhead{} &
    \colhead{${\rm s^{-1}}$} &
    \colhead{$10^{-11}{\rm s^{-2}}$}&
    \colhead{$10^{-21}{\rm s^{-3}}$}&
    \colhead{MJD}
}
\startdata
  B1509-58 & 0020020[108-110]  & 6.582208(4) & - & - & 57936.57452294    \\
  & 0020020[111, 112, 115]  & 6.582218(1)  & -  & - & 57943.05156161  \\
  & 0020020116, 1020020[101,102]  & 6.5820573(9)  & -  & - & 57954.34843163  \\   \hline
  J1811-1925  & 550504[0401, 0601-0605, 0701-1202] & 15.45439494(6) & -1.1070(4)& 8.2(2) & 59769.19550730   \\
    \hline
  J1813-1749$^b$  & 2579030[102-306]  &22.351083818(17)  &-6.4283(22) &  & 58681.04405093       \\
  \hline
   J1838-0655 &251602[0101-0301, 0501-1102]  & 14.18127712(6) & -1.0092(3)& 28.9(2) &  58626.10906066  \\
   & 359802[0301, 0601, 0801, 1001]  & 14.18092139(8) & -1.1032(6) &-344.3(5)  & 59013.35265278  \\
   & 460701[0801-1101]   & 14.1806308(1) & -1.041(2) & -7(2) & 59399.79907331 \\
  \hline
  J1846-0258$^c$ & 251601[0902-11601], 3598010[101-701] & 3.038892369(2) &-6.60460(3) & 5.72(9)& 58893.0     \\
  \hline
  J1849-0001 & 1020660[101-127, 132-134, 137-151, 155-178]  & 25.95900556(4) & -0.9543(2)& 0.98(8) & 58254.80950424  \\
  \hline
  \enddata
  \tablenotetext{\rm a}{The numbers in the $[..]$ represent
    the last three or four digits of IDs  when they share the first seven or six digits.}
  \tablenotetext{\rm b}{Ephemeris taken from \cite{2020MNRAS.498.4396H}.}
  \tablenotetext{\rm c}{Ephemeris taken from \cite{2023ApJ...952..120H}.}
\end{deluxetable*}

\section{Pulse profiles and spectra of the pulsed emissions}
\begin{figure*}
  \epsscale{1}
  \centerline{
    \includegraphics[scale=1]{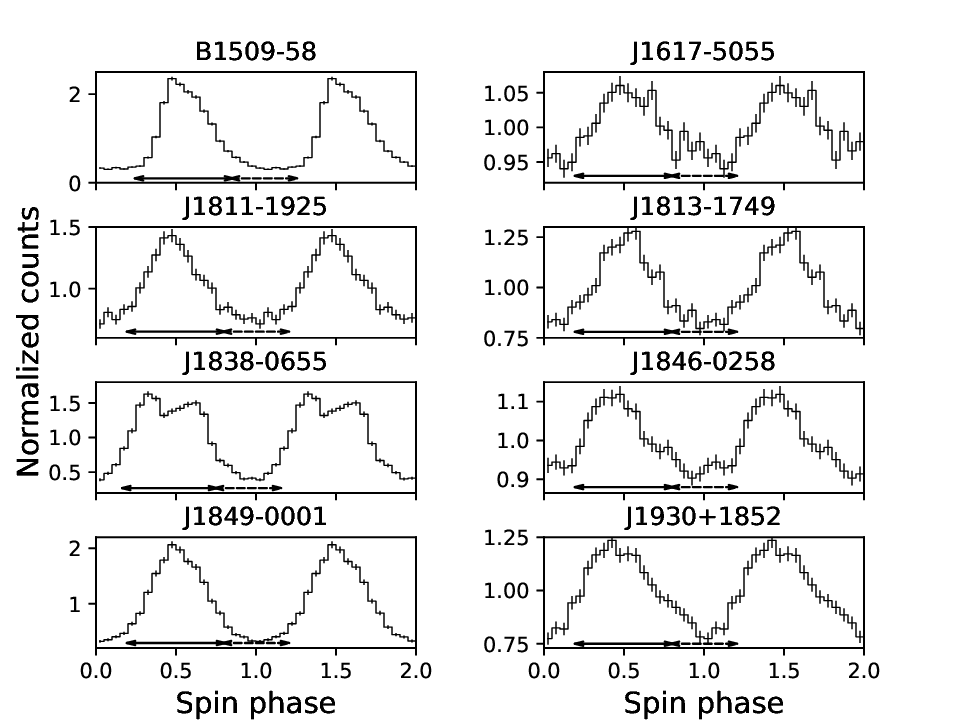}}
  \caption {Pulse profiles of  eight MeV pulsars measure by XMM-Newton. The sold and dashed arrow indicate the on-pulse and off-pulse
  phase intervals, respectively, for the phase-resolved spectroscopy.}
  \label{pulse}
\end{figure*}

\begin{figure*}
  \epsscale{1}
  \centerline{
    \includegraphics[scale=0.6]{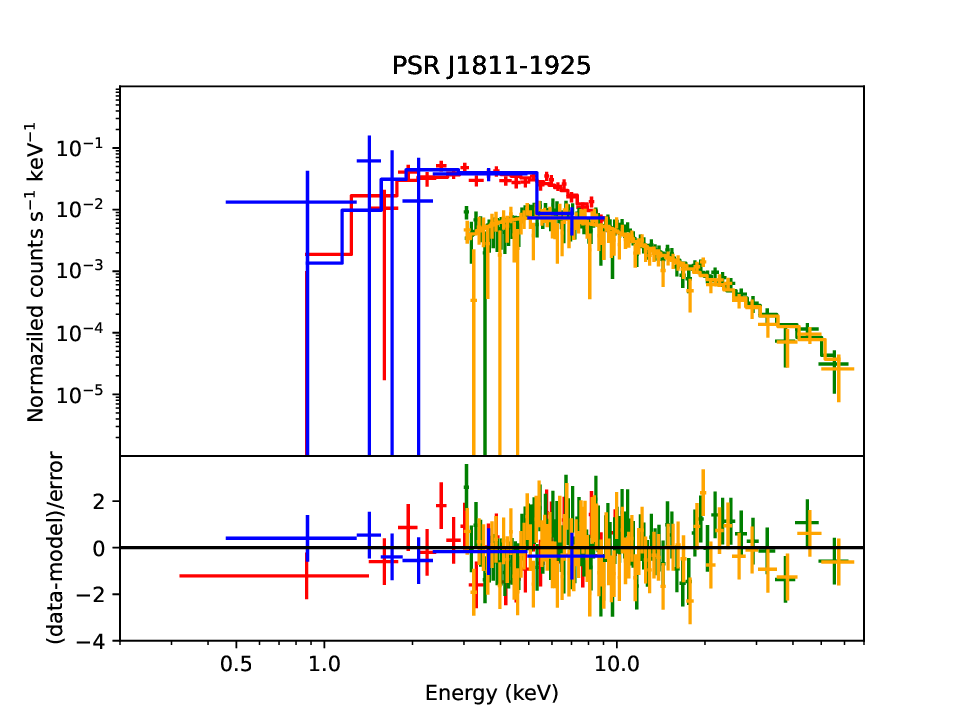}
    \includegraphics[scale=0.6]{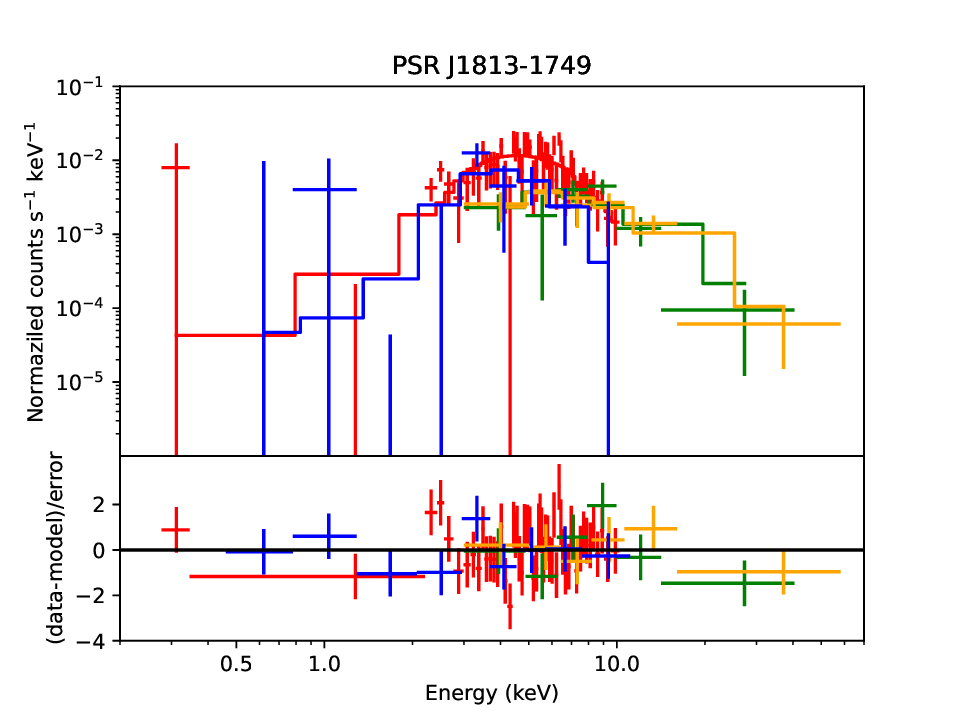}}
  \caption {Same as for Figure~\ref{b1509}, but for PSRs~J1811-1925 (left) and~J1813-1749 (right). }
  \label{j1838}
\end{figure*}

\begin{figure*}
  \epsscale{1}
  \centerline{
    \includegraphics[scale=0.6]{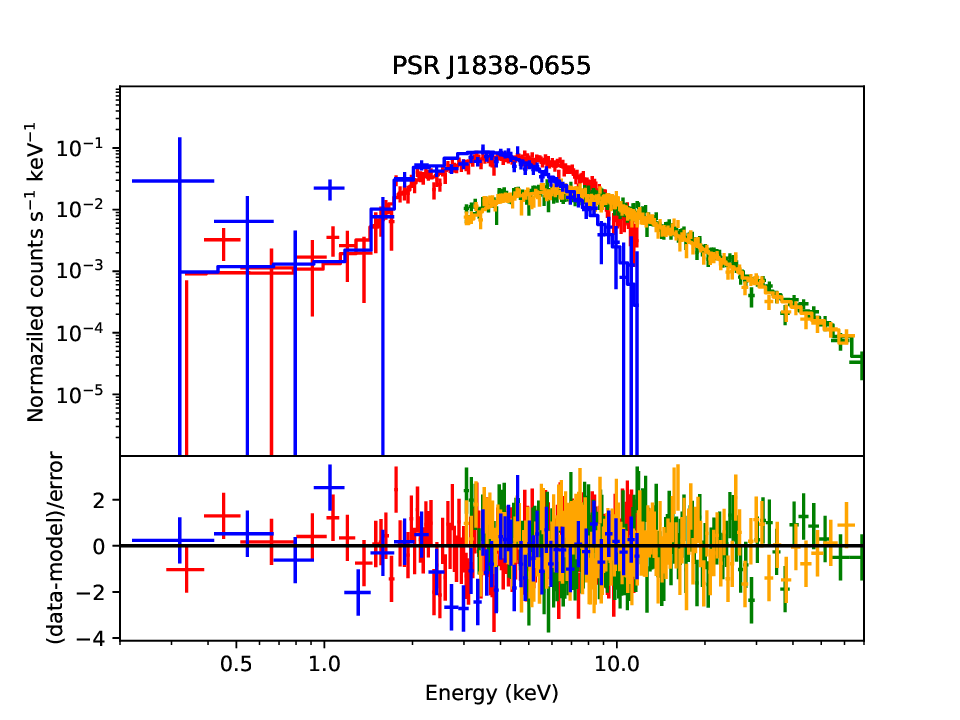}
    \includegraphics[scale=0.6]{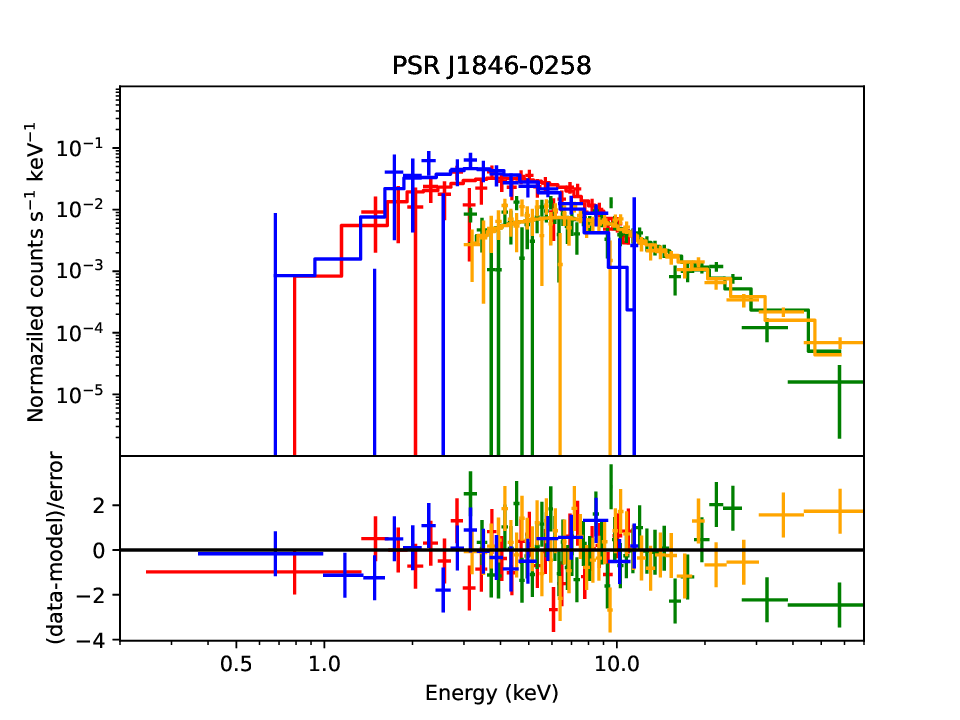}}
  \caption {Same as for Figure~\ref{b1509}, but for PSRs~J1838-0655 (left) and~J1846-0258 (right).}
  \label{j1846}
\end{figure*}

\begin{figure*}
  \epsscale{1}
  \centerline{
    \includegraphics[scale=0.6]{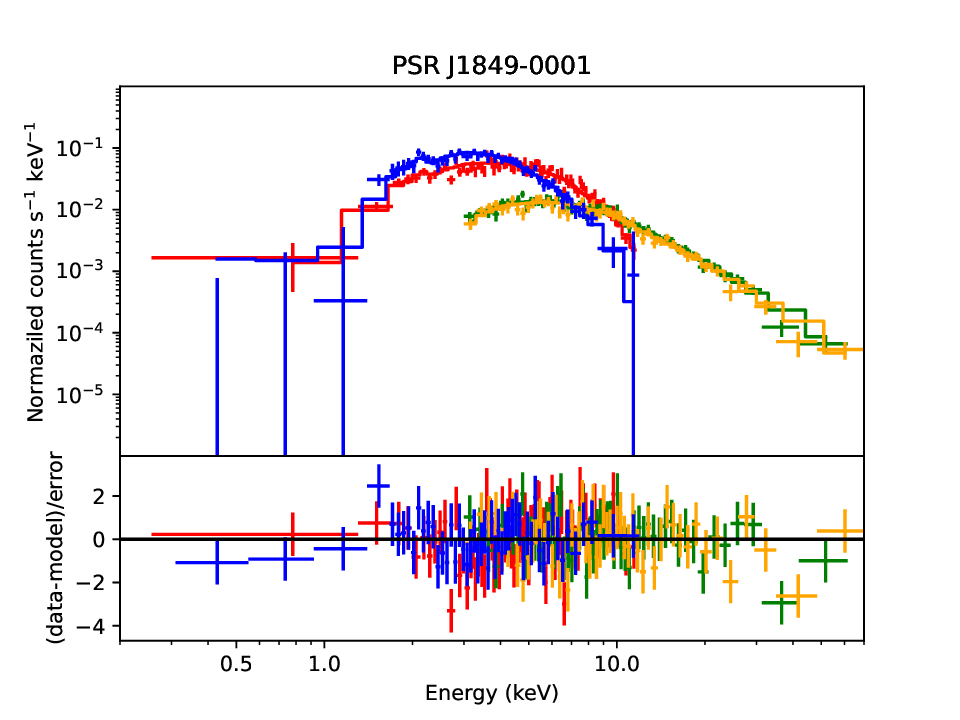}
    \includegraphics[scale=0.6]{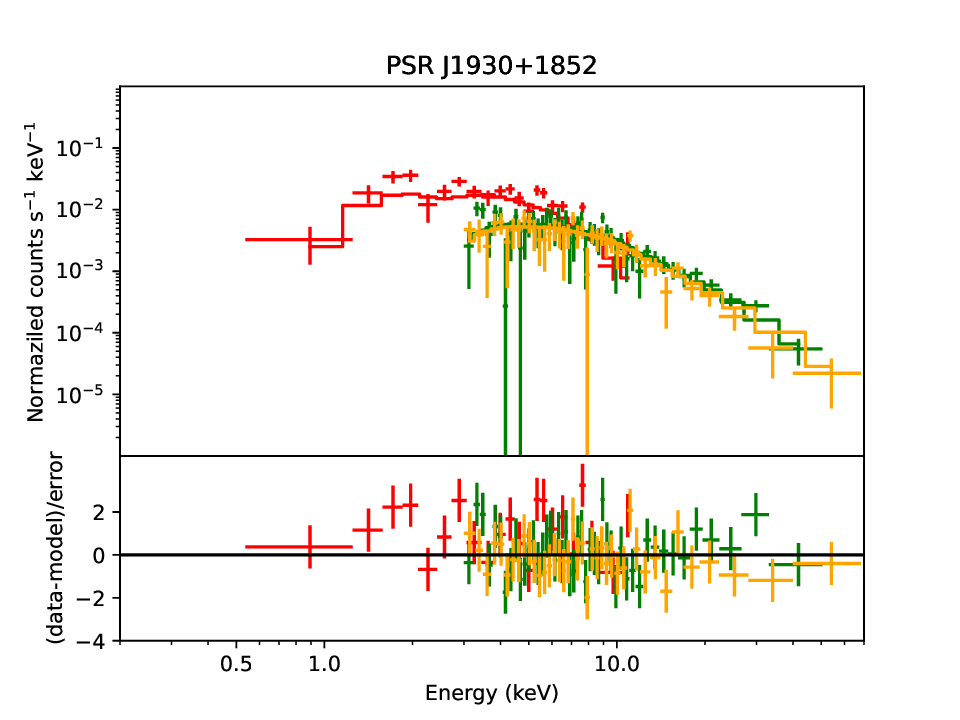}}
  \caption {Same as for Figure~\ref{b1509}, but for PSRs~J1849-0001 (left) and~J1930+1852 (right).}
  \label{j1930}
\end{figure*}

Figure~\ref{pulse} represent the pulse profiles of eight MeV pulsars measured by
the  XMM-Newton observations. The sold and dashed arrows in the figures indicate the  on-pulse and off-pulse phase intervals, respectively, to extract the phase-resolved spectra.   Figures~\ref{j1838}-\ref{j1930} present the spectra of the pulsed emission for PSRs~J1811-1925, J1813-1749, J1838-0655, J1846-0258, J1849-0001 and~J1930+1852, respectively. 

\end{document}